\documentclass{aa} 

  \def\xmm{{\it XMM-Newton~}} 
  \def\swift{{\it Gehrels/Swift~}} 
  \def\chandra{{\it Chandra~}} 
  \def\nustar{{\it NuSTAR~}}

  \usepackage{tablefootnote}
  \usepackage{xcolor}
  \usepackage{natbib}
  \usepackage{graphicx}
  \usepackage{lscape}
  \usepackage{txfonts}
  \usepackage{geometry}
\begin{document}

     \title{Towards a complete census of luminous Compton-thick Active Galactic Nuclei in the local Universe}

      \titlerunning{Complete census of Compton-thick AGN}
      \authorrunning{A. Akylas et al.}
     
     \author{A. Akylas
            \inst{1} 
            \and
            I. Georgantopoulos 
            \inst{1}
            \and
            P. Gandhi
            \inst{2}
            \and 
            P. Boorman
            \inst{3}
            \and
            C. L. Greenwell
            \inst{4}
                        }
     \institute{Institute for Astronomy Astrophysics Space Applications and Remote Sensing (IAASARS), National Observatory of Athens, Ioannou Metaxa \& Vasileos Pavlou, Penteli, 15236, Greece \\ 
     \and
     School of Physics and Astronomy, University of Southampton, Highfield, Southampton, SO17 1BJ, United Kingdom   \\
     \and
     Cahill Center for Astrophysics, California Institute of Technology, 1216 East California Boulevard, Pasadena, CA 91125, USA \\
     \and
     Physics Department, University of Durham, South Road, DH1 3LE, United Kingdom \\
     }

 \abstract
   { X-ray surveys provide the most efficient means for the detection of Active Galactic Nuclei (AGN). However, they face difficulties in
detecting the most heavily obscured Compton-thick AGN.   The BAT detector on board the \swift mission, operating in the very hard 14-195 keV band, has provided the largest  samples of Compton-thick AGN in the local Universe.   However, even these flux limited  samples can miss the most obscured sources among the Compton-thick AGN population. A robust way to find these local sources is  to systematically study volume-limited AGN samples detected in the IR or the optical part of the spectrum. Here, we utilize a local sample  (<100 Mpc) of mid-IR selected AGN, unbiased against obscuration, to determine the fraction of Compton-thick sources in the local universe. When available we acquire X-ray spectral information for the sources in our sample  from previously published studies. Additionally, to maximize the X-ray spectral information for the sources in our sample,  we analyse, for the first time, eleven unexplored  \xmm and \nustar observations, identifying four new  Compton-thick sources. 
Our results reveal an increased fraction of Compton-thick AGN among the sources that have not been detected by BAT of 44 \%.  Overall we estimate a fraction of Compton thick sources in the local universe of 25-30\% among mid-IR selected AGN. We find no evidence for evolution of the AGN Compton-thick fraction with luminosity.

  }


    \authorrunning{Akylas et al.}

    \maketitle
  %

\section{Introduction}
X-ray emission is ubiquitous in Active Galactic Nuclei (AGN). 
This is believed to originate from a corona above the accretion disk
 \citep[e.g.][]{haardt1991}.  According to this picture, the UV photons produced in the accretion disk are scattered by the hot coronal electrons having temperatures of about 60 keV  \citep{Akylas2021} producing copious amounts of X-ray emission. The X-ray radiation that permeates the sky, the X-ray background \cite{Giacconi1962}, 
in the energy range 0.1-300 keV is  produced by the superposition of AGN. 
The {\it Chandra} X-ray mission owing to its superb spatial resolution resolved 
about 90\% of the X-ray background in the relatively soft 0.5-8 keV band \citep{Mushotzky2000}.
Indeed, optical spectroscopic observations confirm that the vast majority of these sources 
are associated with AGN with redshifts peaking at $z\approx 0.7$ \citep[e.g.][]{luo2017}.

 Observations with the \swift/BAT \citep{Ajello2008}, { \it RXTE} \citep{revnivtsev2003}, {\it BeppoSAX} \citep{frontera2007}, {\it Integral} \citep{Churazov2007} missions show that 
the peak energy density of the  X-ray background lies at 
harder energies around 30 keV. 
The X-ray background synthesis models attempt to reproduce the spectrum of the X-ray background by modeling the AGN luminosity function together with their
spectral properties \citep{comastri1995, gilli2007, treister2009, 
akylas2012,ueda2014,ananna2019}.
All X-ray background models find that in order to  reproduce the 30 keV 
energy hump one needs a considerable fraction of heavily obscured Compton-thick AGN. 
However, the exact fraction depends on the amount of reflection in the vicinity of the
emitted radiation that is the accretion disk and the surrounding torus \citep{Gandhi2007,Vasudevan2013}. 
In Compton-thick sources, the line-of-sight to the nucleus 
is obscured by huge amounts of gas with column densities exceeding 
$\rm N_H\approx  10^{24}~cm^{-2}$. 
The obscuring screen is believed to be a dusty structure with toroidal geometry. 
Recent observations suggest that this obscuring screen, the torus, 
is composed of optically thick clouds 
\cite{Nenkova2002, Nenkova2008, Tristram2007}.
 In Compton-thick AGN the obscuration is because 
of Compton-scattering rather than photoelectric absorption. 
 The X-ray photons below 10 keV are almost totally absorbed and hence X-ray observations at higher 
 energies are necessary in order to securely classify a source as a Compton-thick AGN. 
 The exact fraction of Compton-thick sources among the total AGN population differs significantly among the various X-ray background models.
  \cite{akylas2012}  estimate a  rather modest fraction of less than about 20\%, 
  see also \cite{treister2009, Vasudevan2013}. On the other end the models of \cite{ananna2019} find that the Compton-thick AGN could constitute  half of the AGN population. 
  
  The launch of the \swift mission \cite{gehrels2004} made a leap forward in the study of Compton-thick AGN owing to its hard  energy range.  The Burst Alert Telescope, BAT,  onboard \swift continuously 
  scans the whole sky in the 14-195 keV band. 
  Therefore, BAT provided an unprecedented census of the hard X-ray sky and in particular of 
  Compton-thick AGN. \cite{ricci2015}, \cite{akylas2016} analysed the AGN X-ray spectra in the BAT 70-month survey \citep{baumgartner2013} finding a few tens of candidate Compton-thick AGN. 
   In addition, \cite{Marchesi2018}, \cite{zhao2021}, \citet{torres2021} searched for Compton-thick 
   AGN on the Palermo 100-month BAT catalogue. 
   Using the BAT sample \cite{akylas2016} and recently \cite{ananna2022} 
   derived for the first time the Compton-thick AGN  luminosity function in the local Universe. 
   More recently Georgantopoulos et al. (submitted) revisit these luminosity functions using
   the most up-to-date column density estimates derived by  \nustar. 
   According to the above works, the fraction of Compton-thick 
AGN should not exceed 25\% of the total AGN population \citep[see also][]{burlon2011,ricci2015, georgantopoulos2019,torres2021}.

Nevertheless, the Compton-thick AGN detected by BAT may be the tip of the iceberg. This is because 
even the BAT band (14-195 keV) may miss a significant fraction of heavily obscured AGN.
\cite{burlon2011} finds that even mildly Compton-thick AGN,
 with a column density of a few times $\rm 10^{24} cm^{-2}$,
 are attenuated by 50\% in the hard 15-55 keV band. 
The most heavily obscured, reflection dominated Compton-thick 
AGN with column densities $\rm N_H\approx 10^{25}cm^{-2}$ such as NGC 1068 
have the bulk  of their 15-55 keV flux attenuated. 
In order to find the most heavily obscured AGN one has to 
resort to volume-limited optically selected or IR selected AGN.
  Along these lines, 
 \cite{akylas2009} observed with {\it XMM-Newton} all (38) Seyfert-2
 galaxies in the Palomar spectroscopic  sample of galaxies \citep{ho1997a}. 
 They find that the fraction of Compton-thick sources among 
 Seyfert-2 galaxies is about  20\%. 
 Recently, \cite{asmus2020} compiled the most complete so far
 galaxy sample   in
 the local Universe ($\rm <100 Mpc$).  
 They select AGN applying the  selection criteria based on the {\it WISE} 
 W1-W2 colour \citep{assef2018}. This sample comprises of 
 about 150 sources and has been routinely observed by most X-ray missions. For these sources,
 we compile the already published results and we analyse for 
 the first time the X-ray spectra of eleven sources.  Our goal is to provide the most definitive yet estimate
  of the fraction of Compton-thick sources among the
   {\it WISE} selected AGN in the local Universe. 

  Our paper is organised as follows. In section \ref{thesample} we describe in detail the sample of \cite{asmus2020} and the selection criteria  applied. In section \ref{newxray} we detail the new X-ray observations obtained by 
  {\it NuSTAR}, {\it XMM-Newton} and {\it Chandra}. The spectral fit results are presented in Section \ref{SpectralAnalysis}.
  The discussion and the summary are presented in sections 5 and 6 respectively.

\section{The sample}\label{thesample}

In this work we utilize data from the Local AGN survey (LASr) sample, \citep{asmus2020}, which provides the most complete census of mid-IR selected AGN among all galaxies within a volume of 100 Mpc. 
The sample contains 49k galaxies and 
has a completeness of 90\% at $\log M_\star/M_\odot = 9.4$. Then, the  applied {\it WISE} identification criteria 
serve as a robust tracer of AGN emission. 
Briefly, the sources selected satisfy the following criteria: (a) the W1 (3.4$\rm \mu m$), W2($\rm 4.6 \mu m)$ AGN selection criterion from \citep{assef2018}. This criterion is based on the hot emission coming from the AGN heated obscuring torus, becoming prominent 
at short mid-IR wavelengths($\rm <50\mu m$). (b) the theoretical  colour criterion based on the {\it WISE} W2 and W3 ($\rm 12 \mu m$) presented in \cite{satyapal2018}, to separate AGN from luminous starbursts and (c) $\rm L(12\mu m)>10^{42.3}$ 
 $\rm erg~s^{-1}$. According to \cite{asmus2020}, the above criteria ensure a reliability of over 90\% for finding an AGN.

Consequently, the above authors  derive two AGN datasets. The first, is the "known AGN" sample containing {\it WISE} selected AGN, already known to host an active nucleus in the literature. The second, the "new AGN" sample, contains the {\it WISE} selected AGN which have no prior AGN classification in the literature. 

In this study, we primarily focus on the "known AGN" sample.  This is because the "new AGN" sample has practically no X-ray information available. 
On the other hand, as anticipated, the "known AGN" sample contains a large fraction of sources (approximately 75$\%$) detected within \xmm, \nustar  or BAT observations.  The remaining sources (about 25\%) are not present in the above archives  and lack usable X-ray data. Among the remaining sources, less than 30\% fall within the footprint of \swift/XRT observations, and even  fewer cases allow for the extraction of a poor quality  X-ray spectrum. 

\begin{figure}
\begin{center}
\includegraphics[width=0.9\columnwidth]{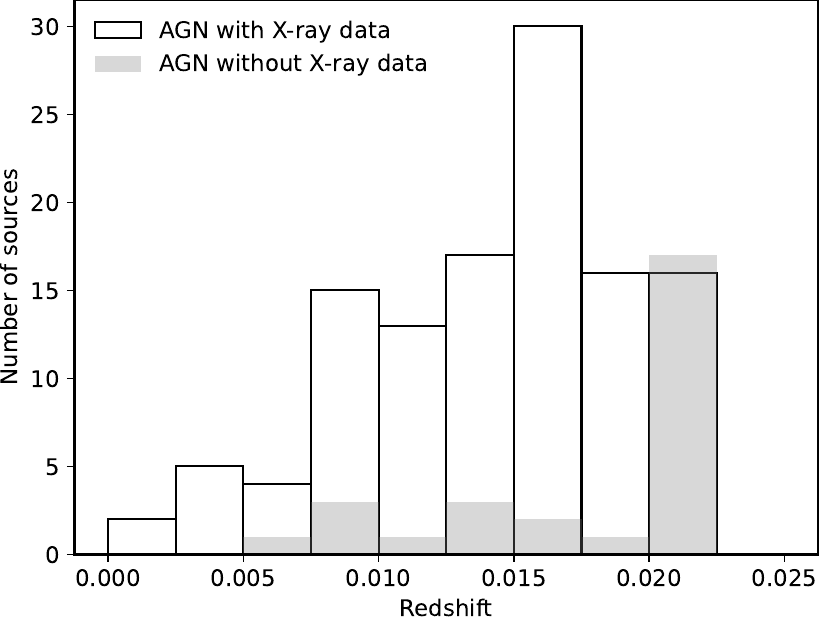}
\end{center}
\caption{The redshift distribution of the sources in the "known AGN" sample. The solid line histogram represents the distribution of sources with available  X-ray data  and the gray shaded histogram the distribution of the sources without X-ray data. }
\label{zcut}
\end{figure}

In Fig. \ref{zcut} we compare the redshift distribution of the sources within these two sub-samples. Notably, the majority of the sources lacking X-ray data are concentrated in the highest redshift bin, i.e.  between 0.02 and 0.022. At lower redshift values the fraction of the sources without X-ray data is very small. Based on this plot and in order to improve the X-ray completeness of our sample, we apply a cut at the redshift  of z=0.02 ($\sim87$ Mpc).

After applying all the selection criteria to the optically selected, "known AGN sample", we have identified  113 AGN. Among these, the largest number  (72 sources) have also been detected in the 70-month Swift/BAT all-sky survey \citep{baumgartner2013}. Their X-ray spectral properties have been systematically studied in detail in \citet{ricci2017} primarily using  \xmm and \swift/BAT spectra. Additionally,  several individual studies re-analysed specific cases, particularly those considered as Compton-thick candidates, utilizing also \nustar observations. 

Among the  41 sources that have not been detected by BAT, 14   have already been analysed using archival \nustar and/or \xmm observations and their results have been  presented in the literature. Additionally a search within the \nustar and \xmm archives
yields good quality X-ray data for eleven additional sources which have not been reported in the literature.
For the remaining sources, those lacking \nustar \xmm or BAT observations, we query the {\it Gehrels/swift} X-ray Telescope (XRT) Science Data Centre repository \citep{evans2009} to obtain spectral fits, when possible. The cross correlation with the XRT 2SXPS catalogue \citep{evans2020} reveals five sources with spectral information available. Thus only eleven sources (i.e less than 10\%) of our sample remains without X-ray spectral information.  After the above selections  our working sample with available X-ray observation is composed of 102 sources
 (see Table \ref{numbers}.)
 For the sake of completeness we mention that there are 32 sources in the "new AGN" sample following the same mid-IR selection criteria, up to a redshift of z=0.02. 

 In Table \ref{full_sample} we present the list of our sources.  The X-ray spectral information is obtained from   either from the literature or from our own spectral analysis. 
 A detailed discussion of the newly analyzed sources is presented in 
 sections \ref{newxray} and \ref{SpectralAnalysis}.

\section{New X-ray observations}\label{newxray}

In Table \ref{thenewsample} we present the X-ray log of the eleven sources with available archival \xmm or \nustar X-ray data, which are  presented for the first time in this work.  In particular, five sources have both  \nustar and \xmm data, one source presents only \nustar and \chandra data while five sources  have only \xmm 
observations available. 

In the table there is an additional entry, i.e. the case of IC4995. Recently, \citet{clavijo2022}  reported IC4995 as a Compton-thick  AGN using the same \nustar data set presented here. However, their analysis employed a rather simple model. Given the Compton-thick  nature of this source, we chose to re-analyze the same \nustar data in conjunction  with \xmm observations to derive more accurate  estimations for its luminosity and column density.

In the case of the five sources in our sample with \swift/XRT data,  marked with an $s$ in Table \ref{full_sample}, we have not perform  the X-ray data reduction or analysis. The X-ray spectral data has been retrieved from the \swift/XRT Science Data Centre repository \citep{evans2020}.

\subsection{\xmm data reduction}

We restrict our analysis on \xmm EPIC-pn data as they provide the highest source count rate. 
The observation data files (ODF) and the Pipeline Processing System (PPS) calibrated event lists of each observation are retrieved from the XMM-Newton Science Archive (XSA). These observations are then further processed 
using the \xmm Science Analysis Software (SAS v21.0.0)  \footnote{https://xmm-tools.cosmos.esa.int/external/xmm\_user\_support/\newline documentation/sas\_usg/USG.pdf}. 

All the event files have been re-screened to remove background flares. To do so we create a single event (PATTERN=0), high energy (10-12 keV) light curve, for each observation. We visually inspect all these light curves and determine the threshold count rate (roughly varying between 0.1 and 0.5 count/s) below which the light curve is low and steady. Then we create the good time interval (GTI) file for each observation using {\sc tabgtigen} task. 

In our analysis only single and double events, (i.e., PATTERN<=4 for the EPIC PN) have been used and, in addition, the flag=0 selection expression has been applied to reject events which are close to CCD gaps or bad pixels. 

We define a circular region of 15 arcsec radius for the source area and a 50 arcsec nearby source-free region for the background estimation. Then the source and background files are produced using  {\sc evselect} task and the corresponding auxiliary files using  {\sc rmfgen} and {\sc arfgen} tasks. All the spectra have been grouped to give a minimum of 15 counts per bin. 

\subsection{\nustar data reduction}

The \nustar observations are processed using the data analysis software {\sc NuSTARDAS} v2.1.2 and {\sc CALDB} v.20231017 \footnote{ https://heasarc.gsfc.nasa.gov/docs/nustar/analysis/nustar \newline \_swguide.pdf}.  
We have downloaded the calibrated event list files from the \nustar archive. Inspection of these cleaned event files shows no further affection by flaring events.

Then we extract source and background event files for each of the two \nustar focal plane modules (FPMA \& FPMB) using the {\sc nuproducts} script. 
We  adopt a radius of 60 arcsec for the source spectral extraction, for both FPMA \& FPMB. The background  spectra are extracted from a four times larger (120 arcsec radius) source-free region of the image at an off-axis angle close to the source position. 

\subsection{\chandra data reduction}

During the single \chandra observation presented here, the ACIS was operated in the ACIS-S mode. Our data reduction for the \chandra observation is performed using {\sc CIAO} v.4.15 and the {\sc CALDB} version 4.10.7. 
We use the level 2, fully calibrated events, provided by the \chandra X-Ray Center
standard pipeline process. We extract the spectral and the ancillary files using the {\sc SPECEXTRACT} script in CIAO.  The source spectrum, is extracted from a circular area of 5 arcsec radius  while the background spectrum was extracted from a nearby, source-free circle of 20 arcsec radius. The spectrum is grouped
using the {\sc GRPPHA} task in {\sc ftools} to a minimum number of 15 counts per bin.

\begin{table}
\begin{center}
\caption{"Known AGN" sample}
\begin{tabular}{cc}
Applied criteria & number \\
\hline
 WISE colors & 146 \\
 z<0.02 & 113  \\
X-ray observations & 102 \\
    \end{tabular}
    \label{numbers}
\end{center}
\end{table}

\begin{figure*}
\begin{center}
\includegraphics[width=0.9\columnwidth]{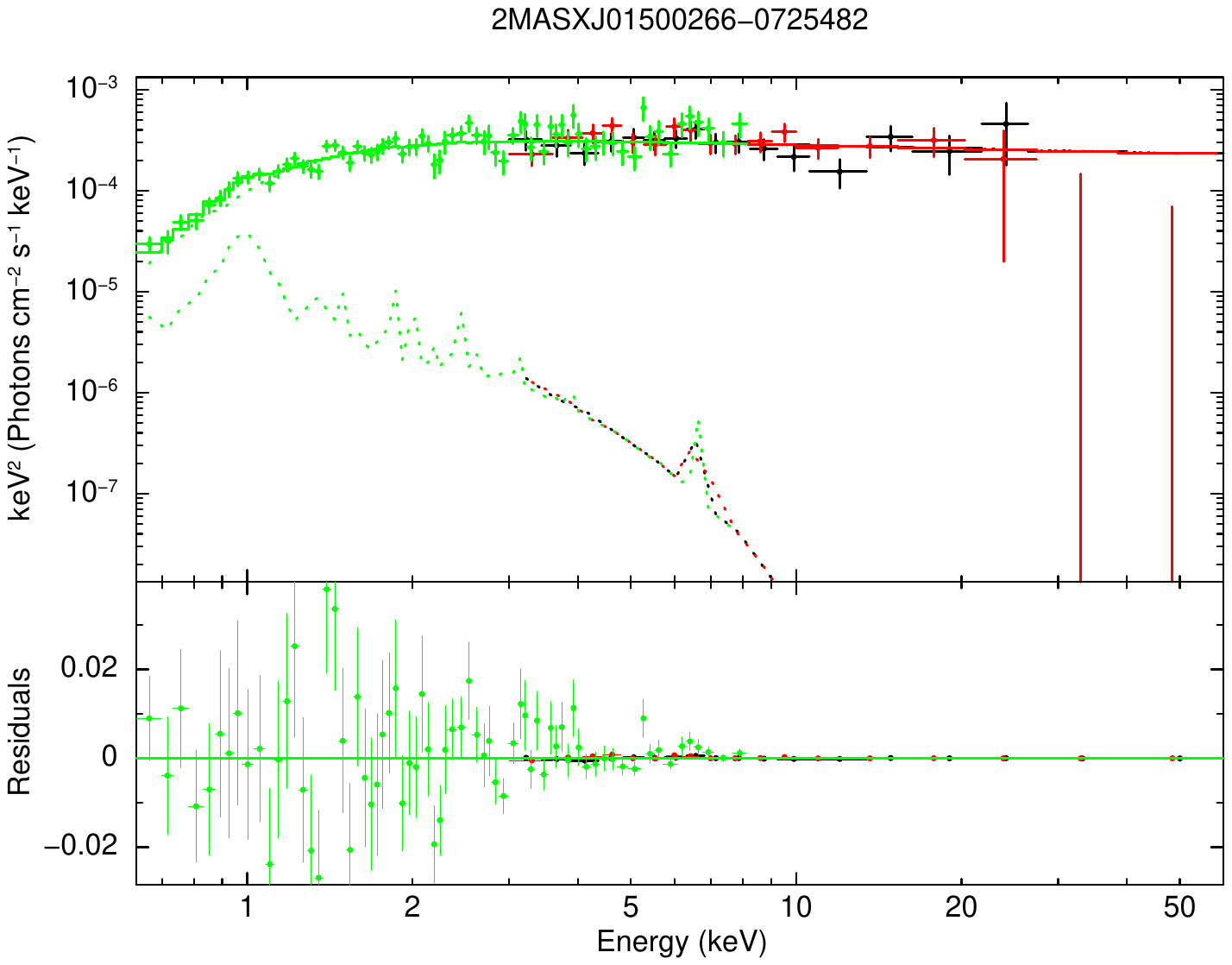}
\includegraphics[width=0.9\columnwidth]{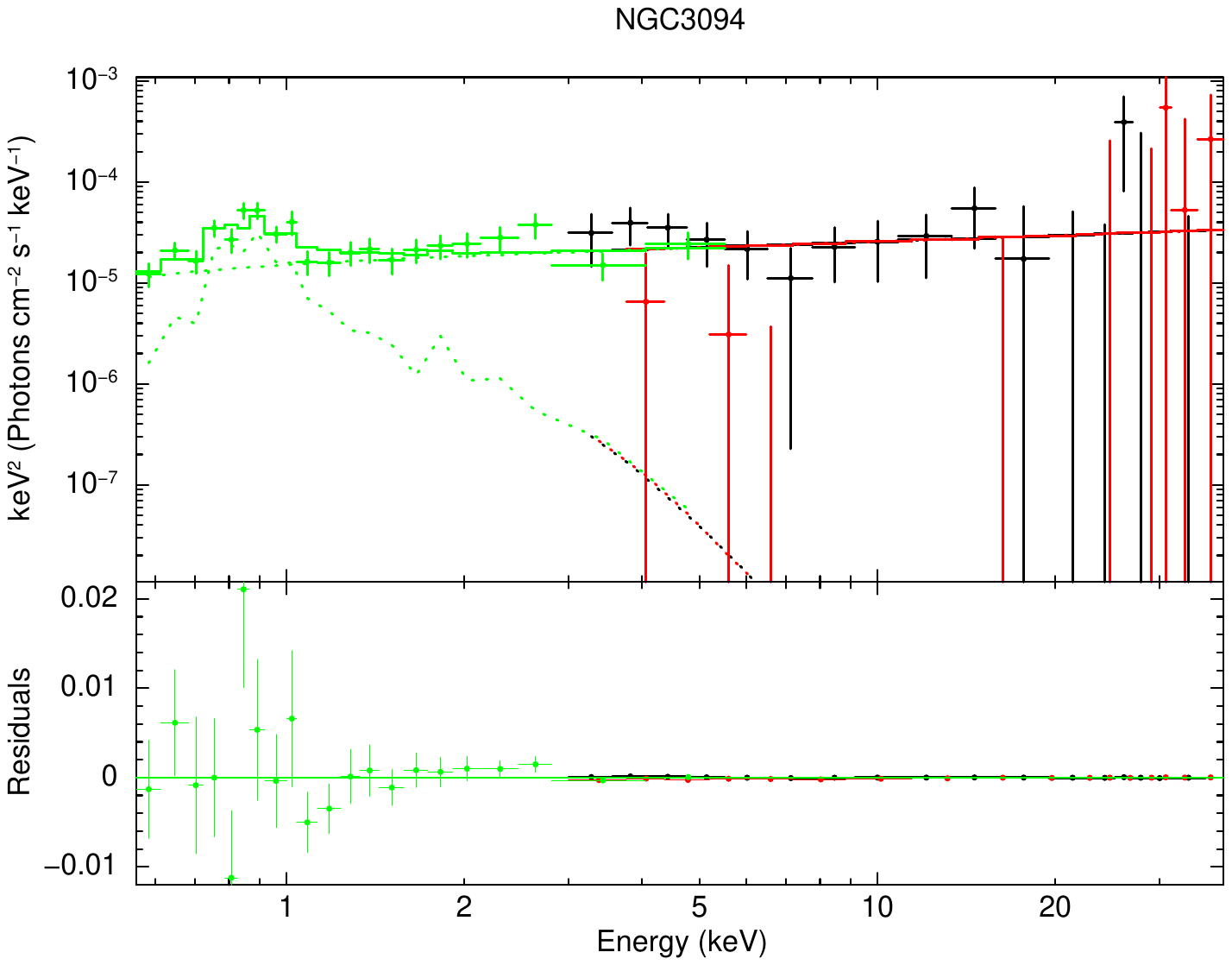}
\includegraphics[width=0.9\columnwidth]{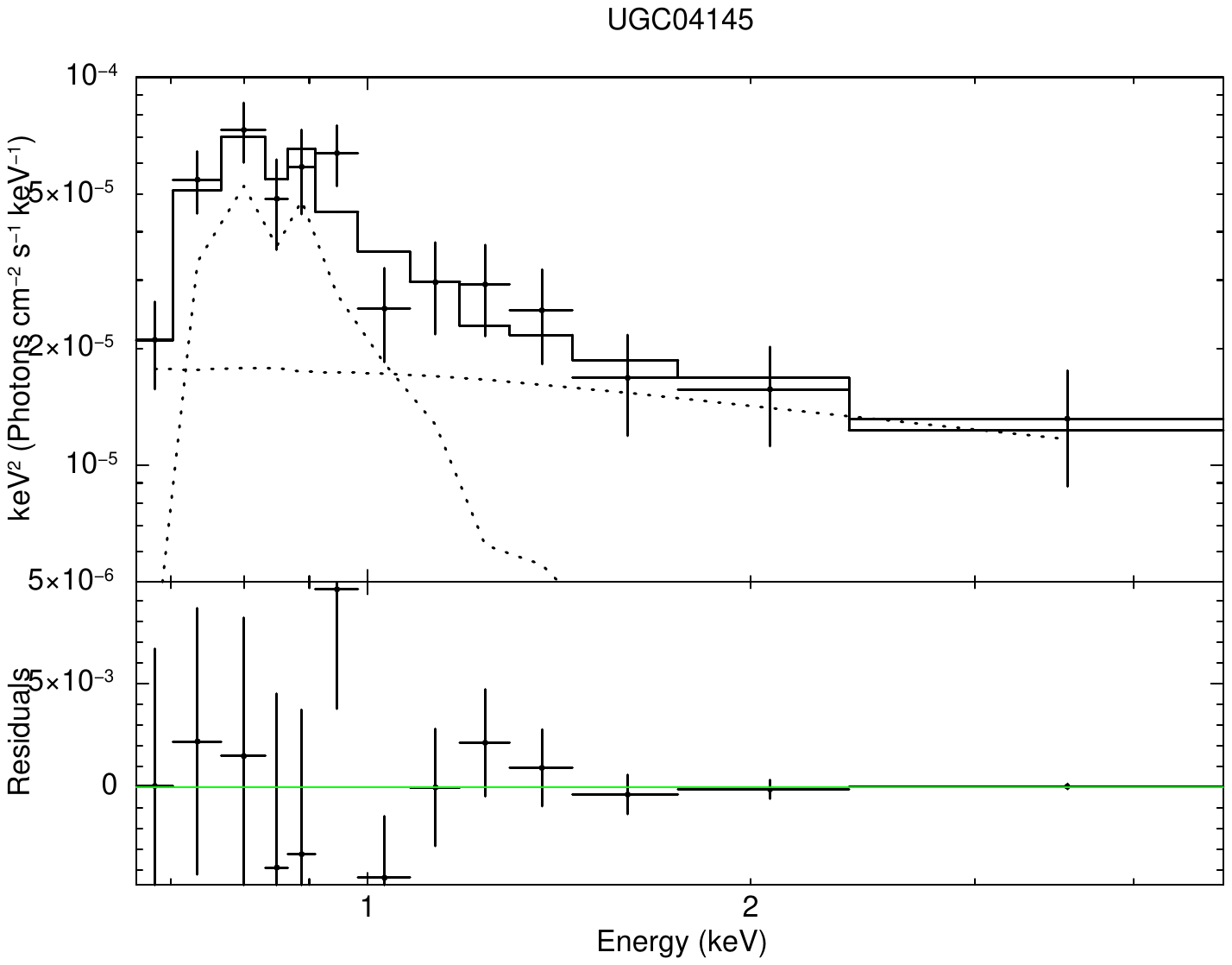}
\end{center}
\caption{X-ray spectra, best fit model and residuals for the sources fitted with the simple model.}  
\label{simple_fit}
\end{figure*}

\begin{figure*}
\begin{center}
\includegraphics[width=0.8\columnwidth]{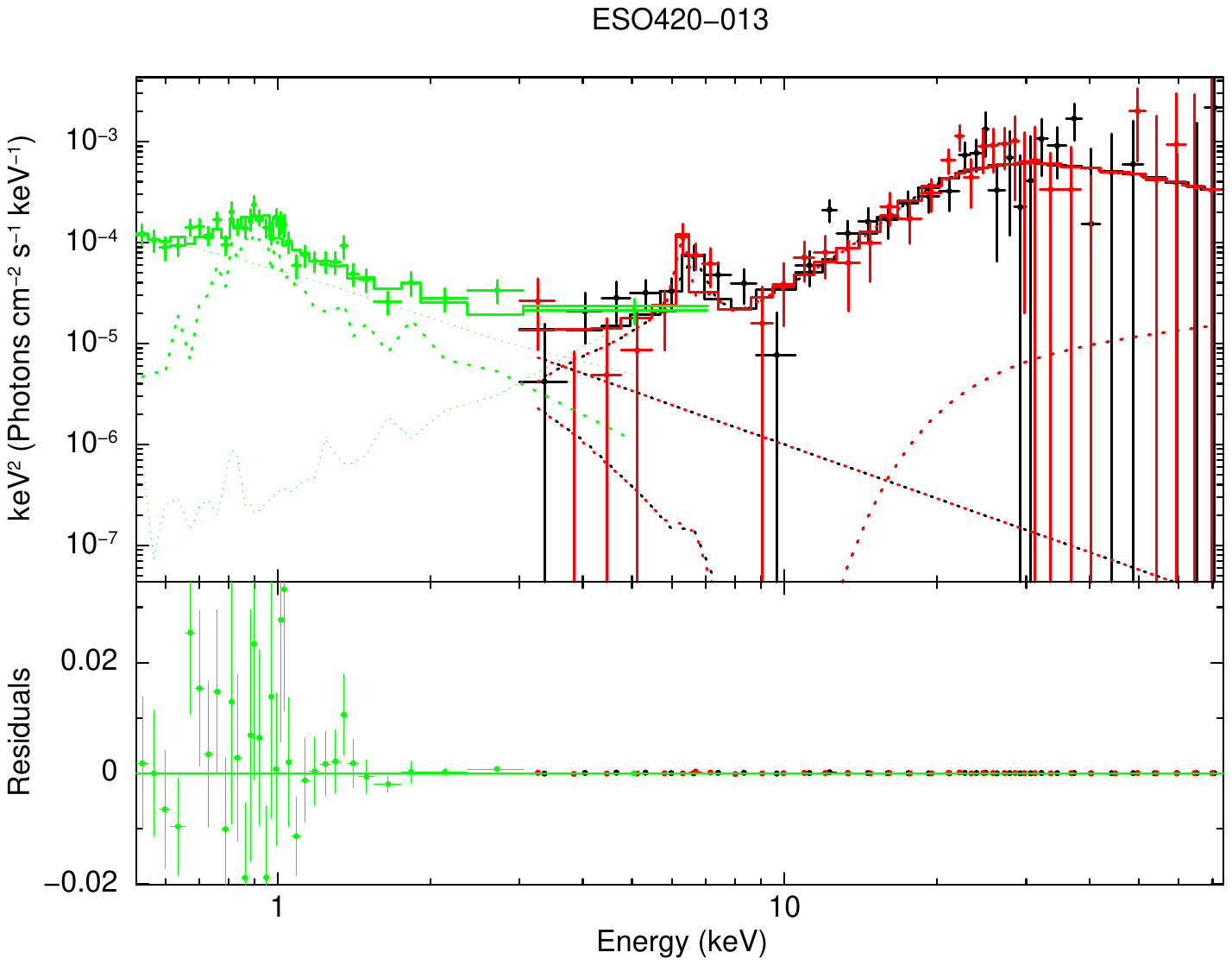}
\includegraphics[width=0.8\columnwidth]{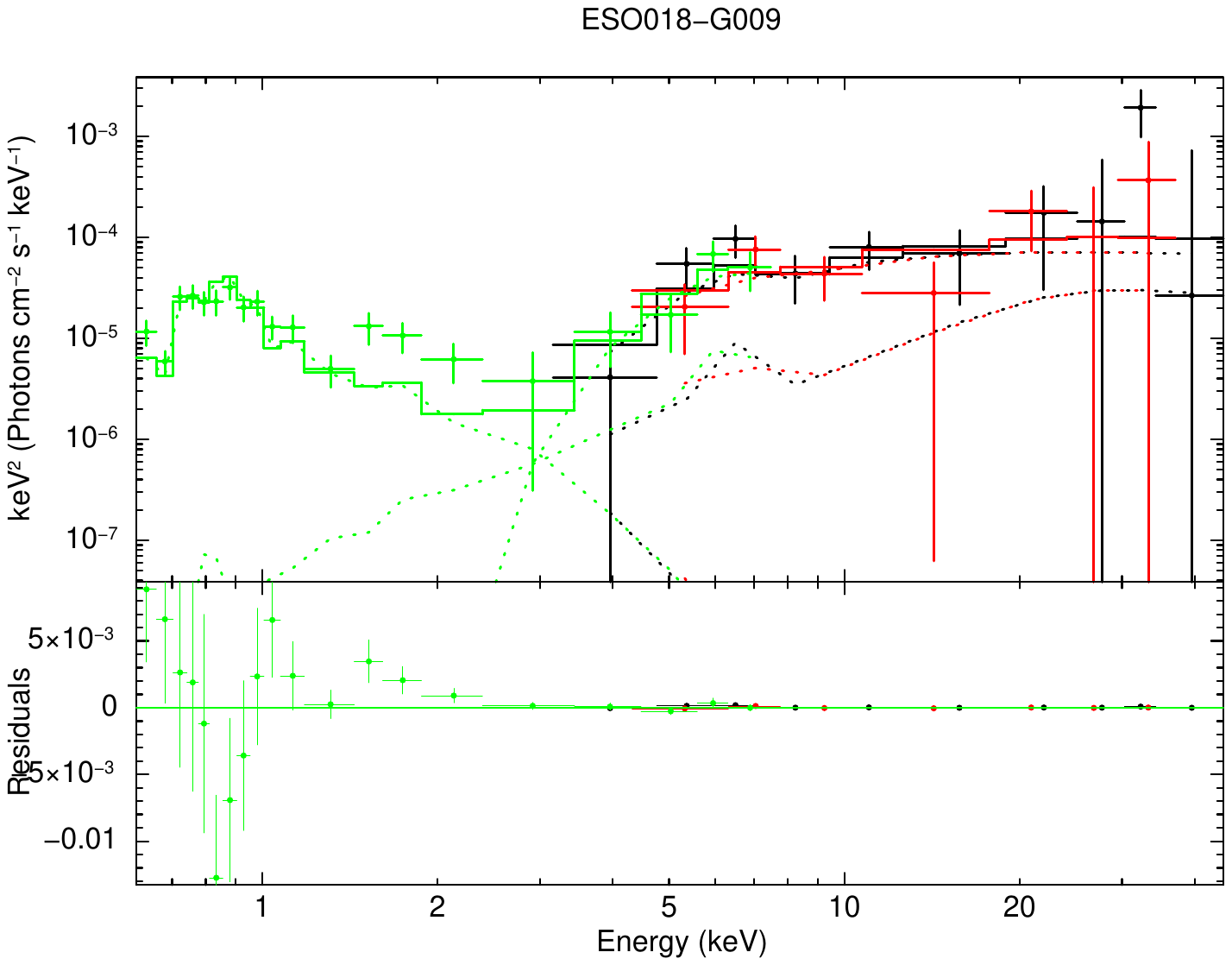}
\includegraphics[width=0.8\columnwidth]{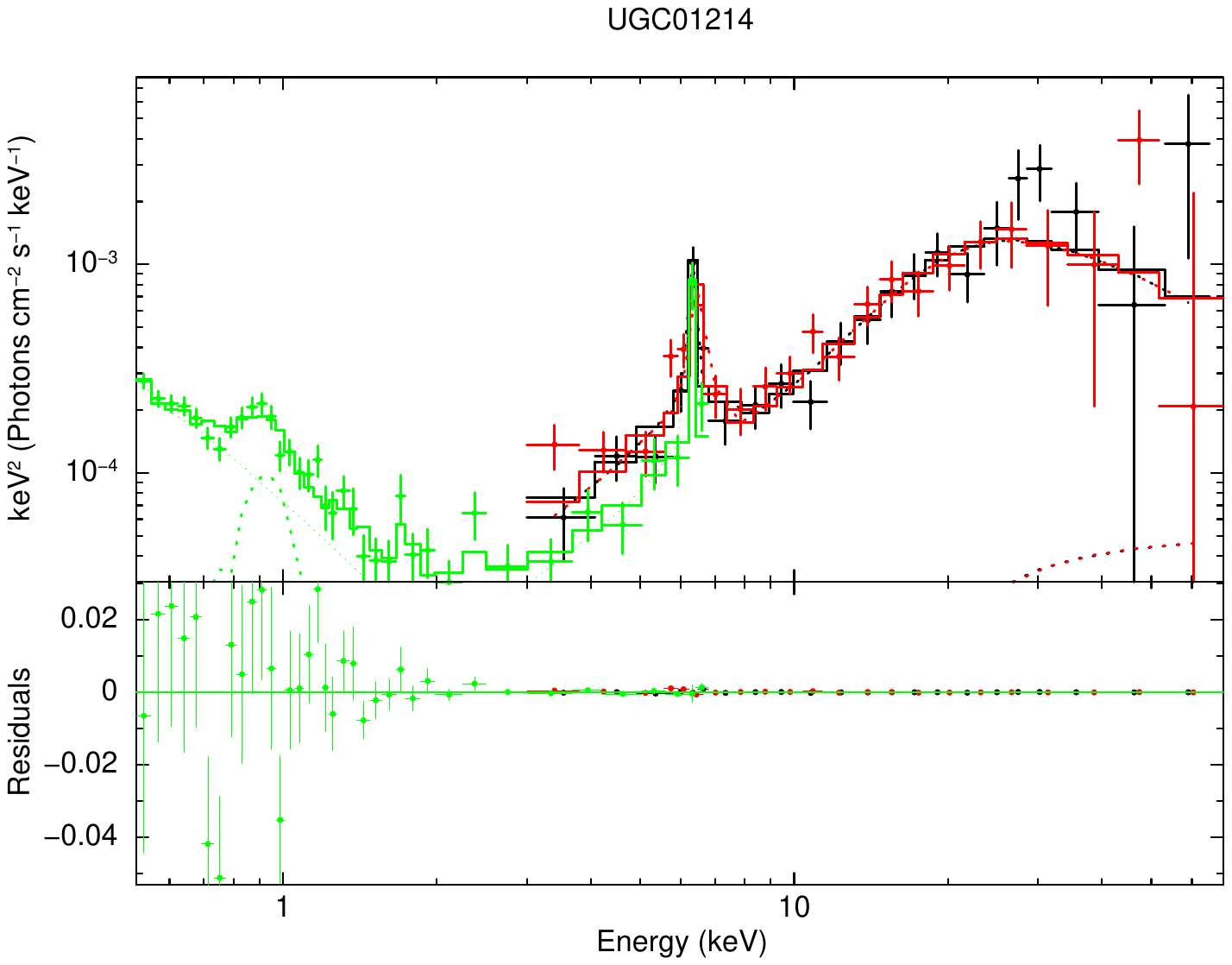}
\includegraphics[width=0.8\columnwidth]{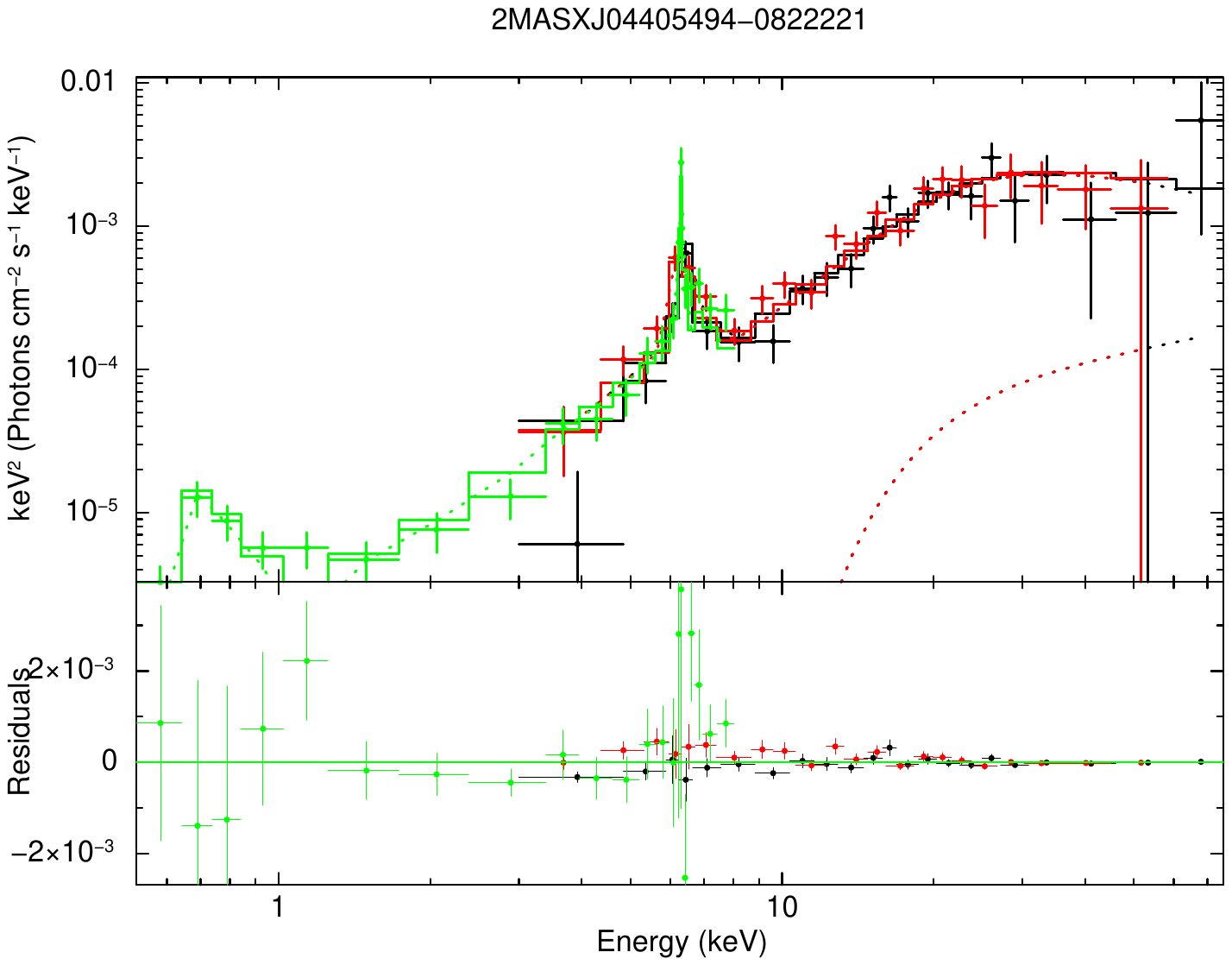}
\includegraphics[width=0.8\columnwidth]{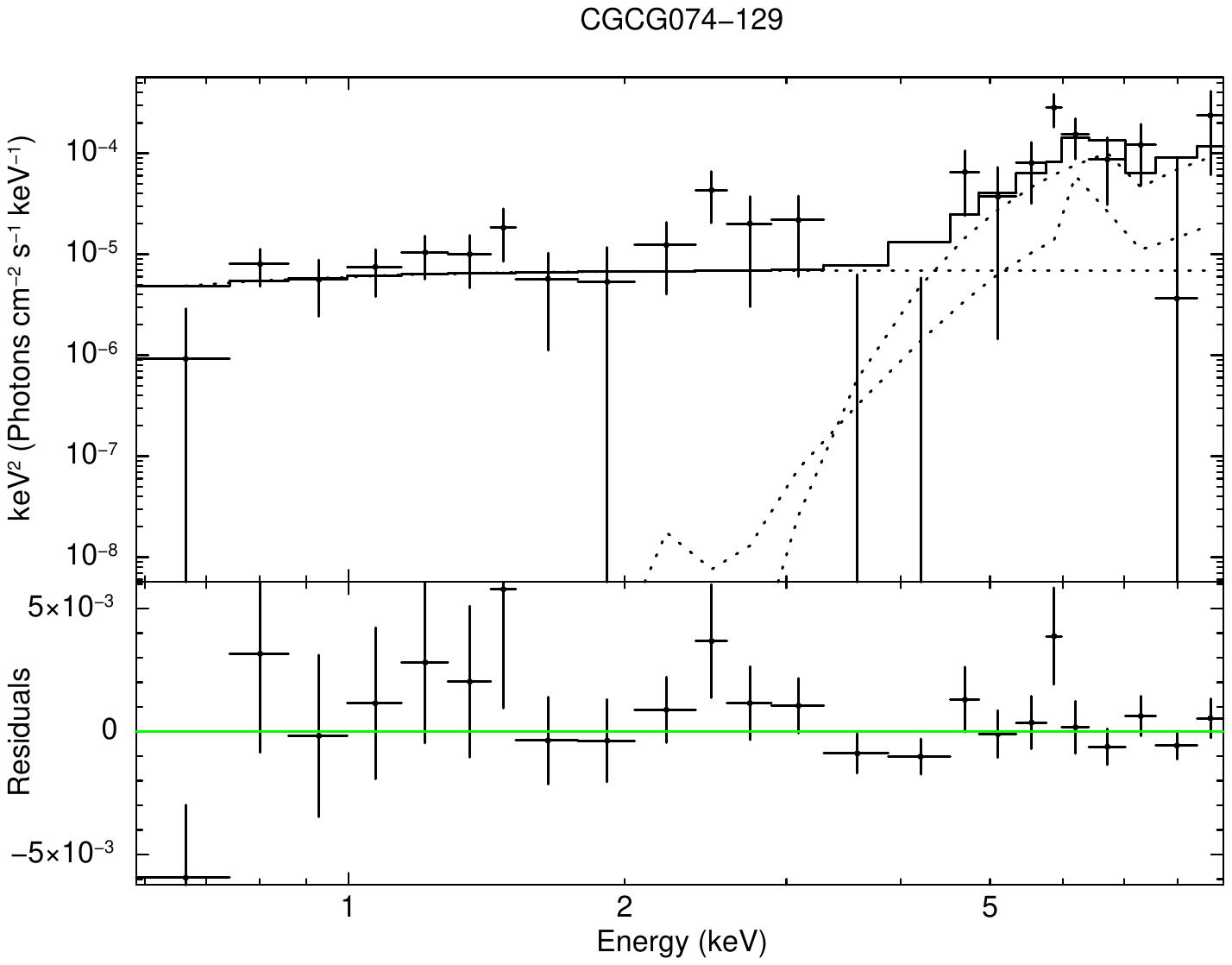}
\includegraphics[width=0.8\columnwidth]{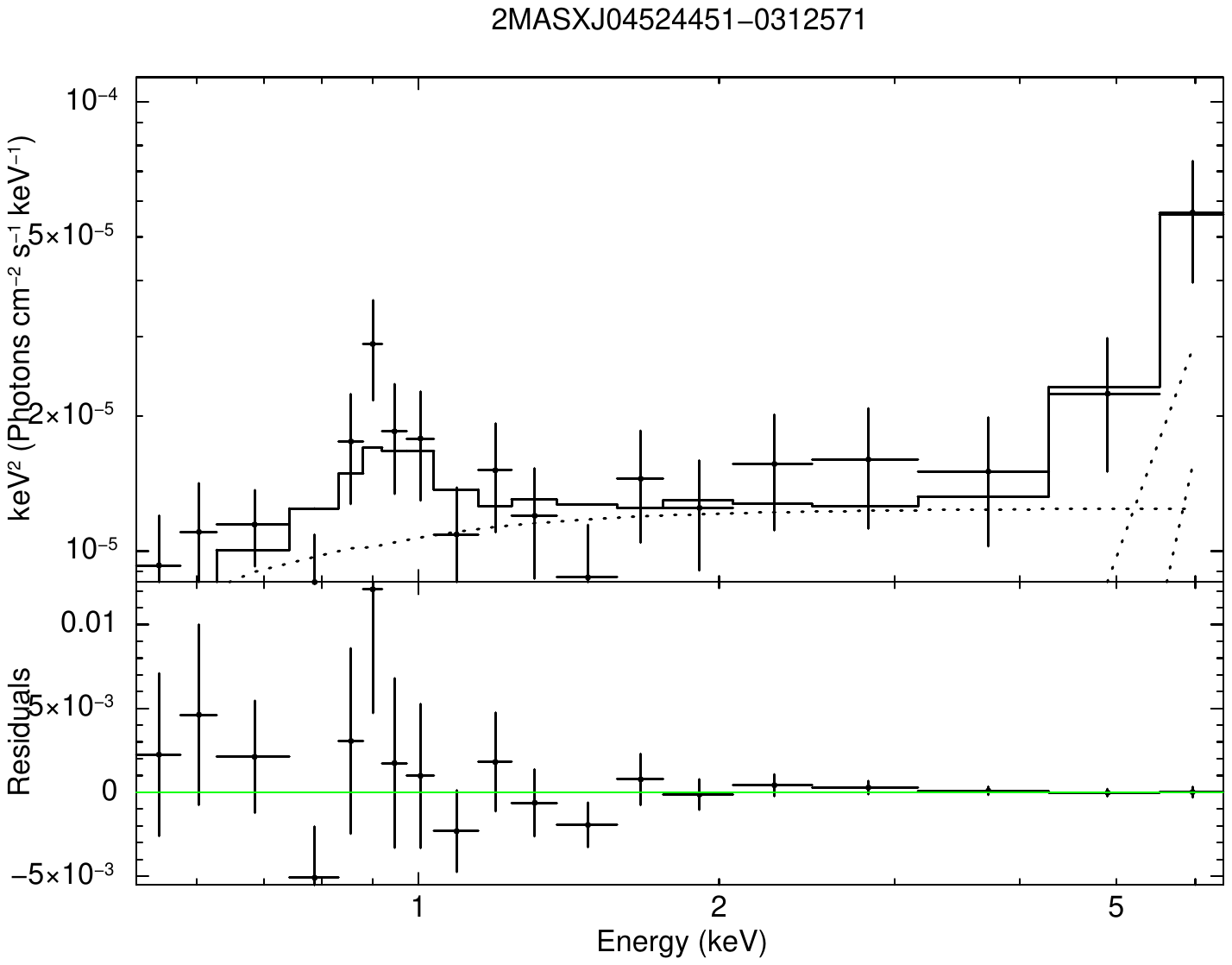}
\includegraphics[width=0.8\columnwidth]{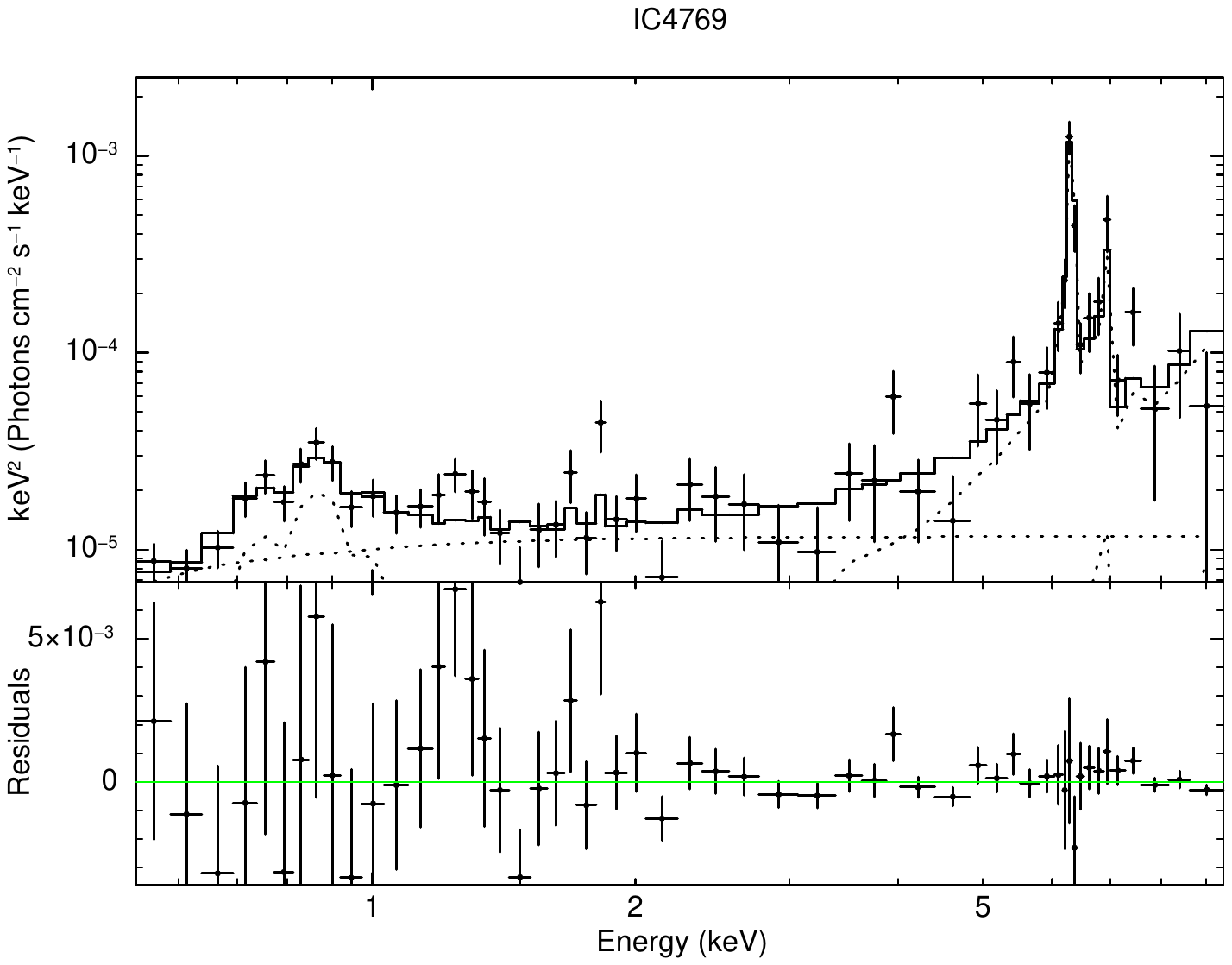}
\includegraphics[width=0.8\columnwidth]{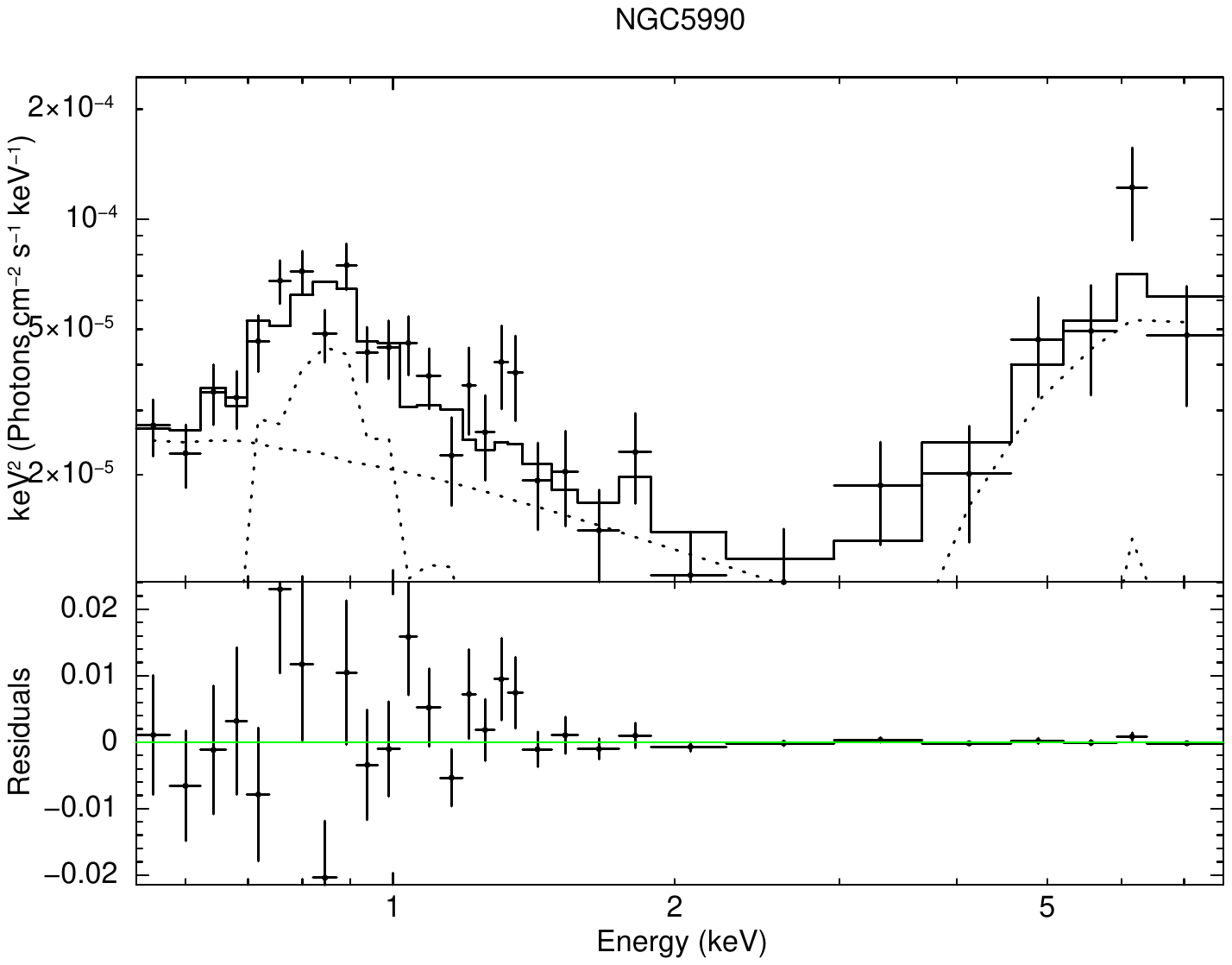}
\includegraphics[width=0.8\columnwidth]{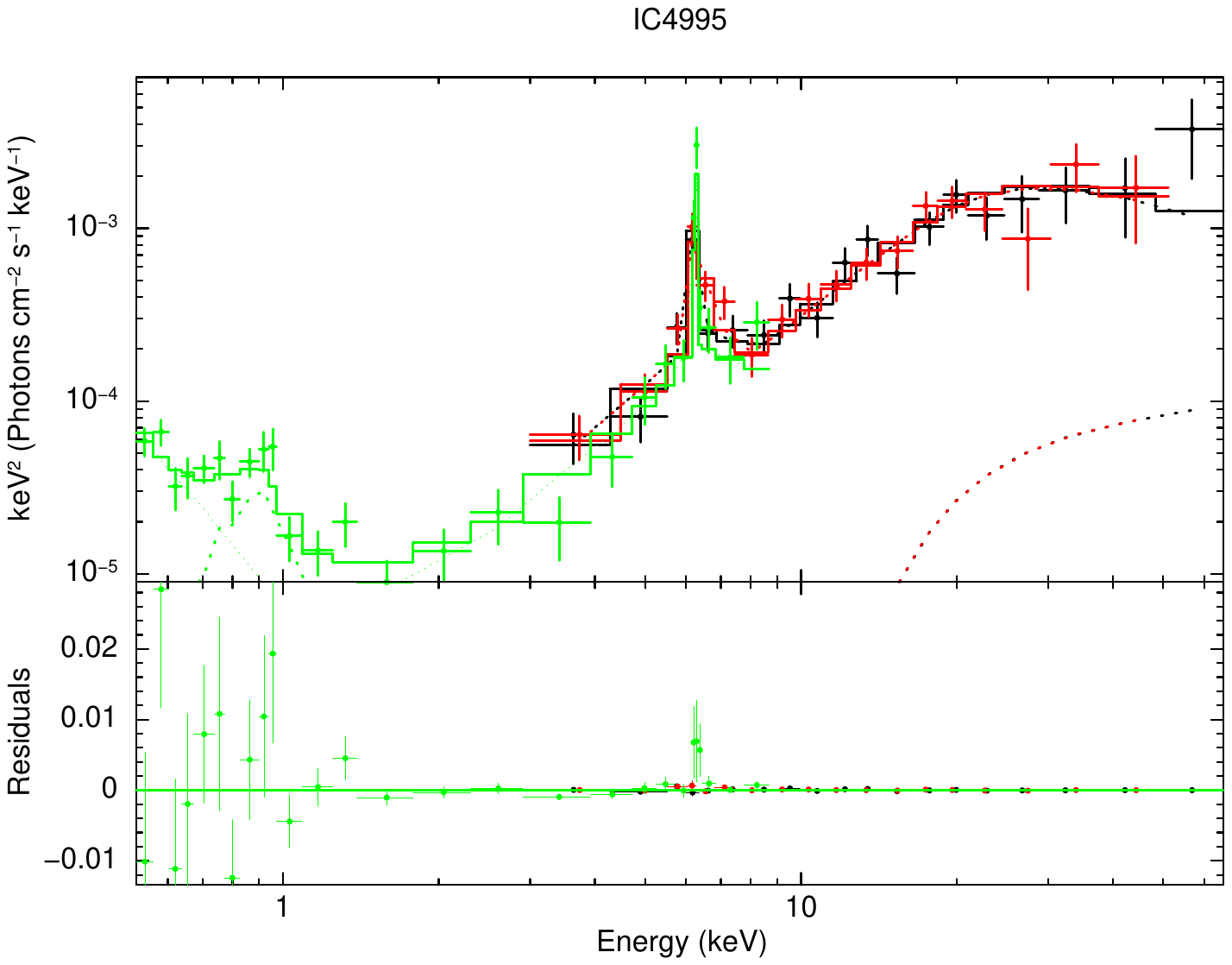}

\end{center}
\caption{X-ray spectra, best fit model and residuals for the sources fitted with the `{\sc RXTORUS} model.}  
\label{advanced_fit}
\end{figure*}

\section{Spectral analysis}\label{SpectralAnalysis}

In this work we present the X-ray spectral analysis for twelve sources. Eleven sources are  presented here for the first time while IC4995,  already presented in \citet{clavijo2022}, has been re-analyzed with a model more suitable 
to its Compton-thick nature.
The spectral fitting was carried out using {\sc XSPEC} v12.13.1
\citep{arnaud1996}. We simultaneously fit all the available spectra for each source.  

\subsection{Single power-law model}
Initially, we utilize a simple model consisting of an absorbed power-law model (PO) to describe the AGN continuum  X-ray emission. We also try to add a Gaussian component (GA) to account for the $\rm Fe K{_\alpha}$ emission line and/or a thermal model (APEC) and/or a second soft power-law to accommodate the possible presence of starforming or scattered soft X-ray emission.  These additional components are added to the data when they provide an improvement to the fit at the 90 per cent confidence level. 

The full {\sc XSPEC} notation of our model is {\sc PHABS$_{GAL}$*(APEC+PO+zPHABS*PO+zGA)}. The Galactic  value of the equivalent hydrogen column for the photoelectric absorption model (PHABS) is being fixed to the value obtained from \citet{dickey1990} while the intrinsic column density of the source is  free to vary. The source redshift is also fixed to the spectroscopic values presented in Table \ref{thesample}. The abundances have been fixed to the default abundance table values in {\sc XSPEC} and the  width of the Gaussian line is also freeze to 0.01 keV.  

Only three sources  are consistent  with this model. The reduced $\rm \chi^2$ which is close to one suggests a good fit. These lack a strong absorption or high $\rm Fe K{\alpha}$ equivalent width.
The spectral fitting results for these three sources are listed in Table \ref{results_simple_fit}. In particular,  we list the value of the plasma temperature for the thermal model, the photon index of the power-law along with the estimated amount of obscuration, the equivalent width of the $\rm Fe K{\alpha}$ line and normalization parameters of the continuum. A goodness of fit estimator is also presented using the $\chi^2$ statistic over the degrees of freedom ($\rm \chi^2$/dof).  All the errors correspond to the 90\% confidence level.
Moreover, Table \ref{fluxes_simple_fit}, lists the estimated flux and luminosity values for each component used in the fitting. In particular, we provide the observed flux of the soft components and the continuum flux in the  0.5-2, 2-10 and 10-80 keV bands. Estimates in the ultra hard 10-80 keV band are presented only when  \nustar data are available. We also present the intrinsic luminosity measurements of the same components by zeroing the value of the intrinsic column density in the best fit model.  

In Fig. \ref{simple_fit} we plot the unfolded spectrum,  the best fit model and the residuals in $\rm E^{2}f(E)$ units. This representation visualises the spectral fitting results and distinguishes between the different spectral components. Note that in an $\rm E^{2}f(E)$ plot a photon index of $\Gamma=2$ is represented by a horizontal line. The thermal emission is shown as an increase above the hard X-ray continuum in the lower energies, while absorption effects appear as a constant decline of the spectrum towards lower energy part of the spectrum.  

\subsection{RXTORUS model}
For the  majority of the sources, the spectral fitting results using the above simple model do not provide an acceptable fit. This is primarily due to the estimated column density being sufficiently high, close to or exceeding the Compton-thick limit. Consequently,  Thompson scattering effects must also be considered through  appropriate modeling. Moreover, in these cases, reflected emission may be a dominant process in shaping of the hard X-ray spectrum. If not properly addressed, this can result in unrealistic X-ray spectral indexes and inaccurate fitting outcomes. 

To address these complexities we opt to utilize the {\sc RXTORUS} model introduced by \citet{paltani2017}. This model self-consistently addresses  (a) the transmitted continuum, containing only photons that did not 
interact with the surrounding material, (b) the scattered continuum, containing  all photons that underwent one or more Compton scatterings, but no photo-ionization or fluorescence and (c) all photons that underwent at least one photo-ionization and subsequent fluorescence re-emission, in addition to any number of Compton scatterings.

The free model parameters are  (a) the power-law photon index, (b) the column density along the line-of-sight, (c) the equatorial column density of the torus, (c) the inclination angle, defined as the angle between the normal to the plane of the torus and the observer, (d) the opening angle of the torus defined as the ratio of the inner to the outer radius of the torus and (e) the normalization factor. 
In our case we use  the {\sc RXTORUS} model grid calculated with a pre-fixed high energy cut-off value of 200 keV,  and the abundances are fixed to the default abundance table values used in the {\sc XSPEC} \citep{anders1989}. In our set-up we assume a simple torus geometry and therefore the equatorial column density and the line-of-sight obscuration are tied using Eq. 1 in \citet{paltani2017}. Also, the normalization parameters of the reprocessed emission are tied to the normalization of the continuum. 

As in the case of the simple modeling presenting above, we check for the presence of a thermal model (APEC) and/or a second soft power-law to account for any starforming or scattered soft X-ray emission.  These additional components are added to the data when they provide and improvement to the fit at a 90 per cent confidence level at least. 
The  {\sc XSPEC} notation of this model is $\rm PHABS_{GAL}*(PO+APEC+RXTORUS)$, where the {\sc RXTORUS} notation corresponds to the {\sc XSPEC} command atable(RXTORUS reprocessed component)+etable(RXTORUS continuum component)*CUTOFFPL.

In specific cases, depending on the quality of the data and model complexity, we choose to freeze certain model parameters to aid in  fitting and prevent convergence issues. Thus, the photon index is sometimes fixed at the value of $\Gamma=2$, the viewing angle is set to 90 degrees (edge-on) and the ratio of the inner to the outer radius is fixed at 0.5, corresponding to a half torus opening angle of 60 degrees.

The spectral fitting results using the above spectral model are presented in Table \ref{results_advanced_model}. In detail,  we list the  thermal model plasma temperature,  the photon index of the scattered emission ($\rm \Gamma_{soft}$), the photon index of the continuum power-law along, the  obscuration along the line of sight, and the corresponding equatorial column density. The values for the torus opening angle and the viewing angle are also listed in the same table. An estimate of  the goodness-of-fit is also provided using $\chi^2$/d.o.f. ratio.  Model components that are not statistically significant are omitted from the table. Fixed parameter values and upper limit estimations are clearly indicated in the table. All  errors correspond to the 90\% confidence interval. Notably, all the sources are reasonably well fit using the {\sc RXTORUS} model.
Moreover, in Table \ref{fluxes_advanced_fit}, we provide a comprehensive summary of the estimated flux and luminosity for each of the  model components used in the fitting as discussed previously in the case of the simple spectral modeling.  Additionally,  in Fig. \ref{advanced_fit} we plot the unfolded spectrum, the best fit model and the residuals in $\rm E^{2}f(E)$ units.

Our spectral analysis suggests that all sources present column densities exceeding $\rm 10^{23} cm^{-2}$.
In particular, there are four Compton-thick sources, reported for the first time in the literature: ESO420-013, UGC01214, 2MASXJ04405494-0822221 and IC4769. In three cases we can only provide a lower limit  on $\rm N_H$ estimation utilizing both \nustar ~ and \xmm ~ data.  Only in the case of IC4769 the $\rm N_H$ constrain is tighter, achieved however by fitting the only available {\it XMM-Newton} data with a fixed photon index. 
A common characteristic is the very prominent Fe$\rm K\alpha$ emission line, present in all the individual observations. This further supports the presence of high amounts of obscuring material along their line of sight. 
 
In all sources strong soft excess emission is being observed,  originating from optically thin collisionally ionized hot plasma ({\sc APEC} model component). Additionally, 
most sources require an additional scattered emission component (soft power-law component) to adequately fit the soft X-ray data.  When the data do not provide a reasonable constraint, the photon indices of both the soft and hard power-law components are tied and fixed to a value of $\Gamma=2$. 
However, in certain cases, decoupling the photon index of the scattered emission from that of the intrinsic emission as suggested by,  in \cite{silver2022} and \cite{yamada2021}) leads to a significantly improved spectral fits. 

\subsection{UXCLUMPY model}
For the five Compton-thick sources presented in Table \ref{results_advanced_model} we repeat the spectral fitting analysis using the {\sc UXCLUMPY} model \citep{buchner2019}. 
This model is constructed to reproduce the column density distribution of the AGN population and cloud eclipse events in terms of their angular sizes and frequency. The model assumes a clumpy structure for the obscuring material and applies a second Compton thick reflecting layer close to the corona. 
Our motivation is to verify the Compton-thick nature of these sources using a substantially different fitting model. An equally important task is to explore whether the {\sc UXCLUMPY} model, allowing for much higher values for the column density than any other model,  up to the $10^{26} \rm cm^{-2}$, could provide tighter  constrains for the column density of these sources. To this end we use a similar setup as described previously; however all the geometrical parameters have been kept fixed. The  viewing angle, measured from the vertical symmetry axis toward the equator has been fixed to 90 degrees. The $\sigma_{torus}$ parameter  (TORsigma in {\sl XSPEC} table model) that describes the dispersion of the cloud population has been fixed to 30 degrees,  and the covering factor of the Compton thick inner reflecting layer, (CTKcover parameter), has been fixed to 0.4 . The results are presented in Table \ref{results_uxclumpy_model}. The new spectral fitting results  verify the Compton-thick  nature of these sources. However, we are still unable to  further constrain the $\rm N_H$ values which still appear as lower limits.

\begin{figure}
\begin{center}
\includegraphics[width=0.9\columnwidth]{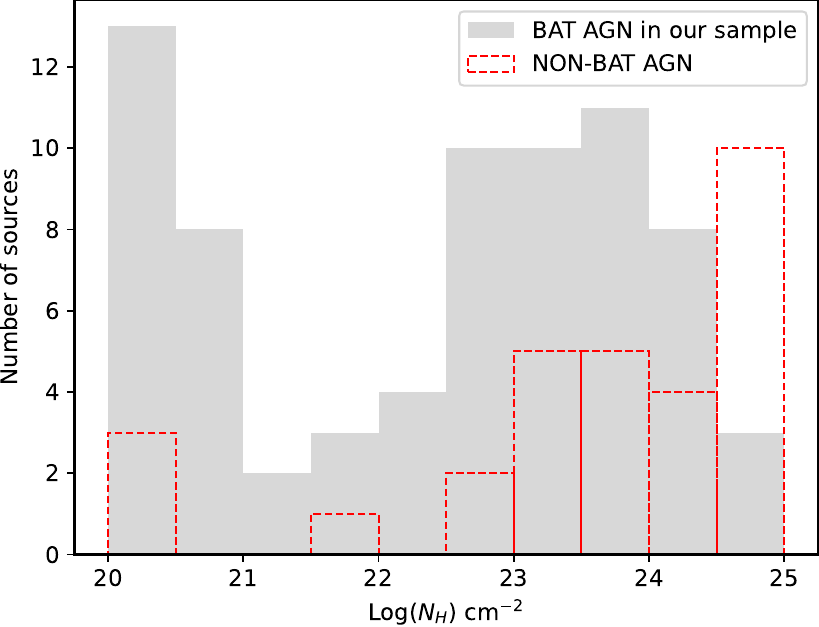}
\includegraphics[width=0.9\columnwidth]{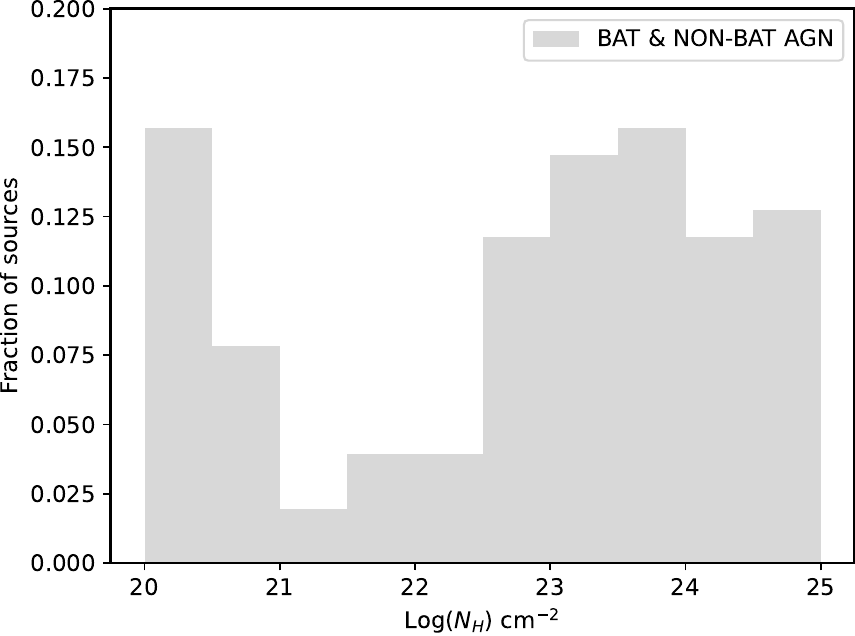}
\end{center}
\caption{(a) Upper panel: Comparison of the $\rm N_H$ distribution for the sources in our sample detected in \swift/BAT 70 months all sky survey (grey-shaded histogram) with  those missed (red-dashed line). (b) Lower panel: The $\rm N_H$ distribution for the total population}  
\label{nh_dist}
\end{figure}

\begin{figure}
\begin{center}
\includegraphics[width=0.9\columnwidth]{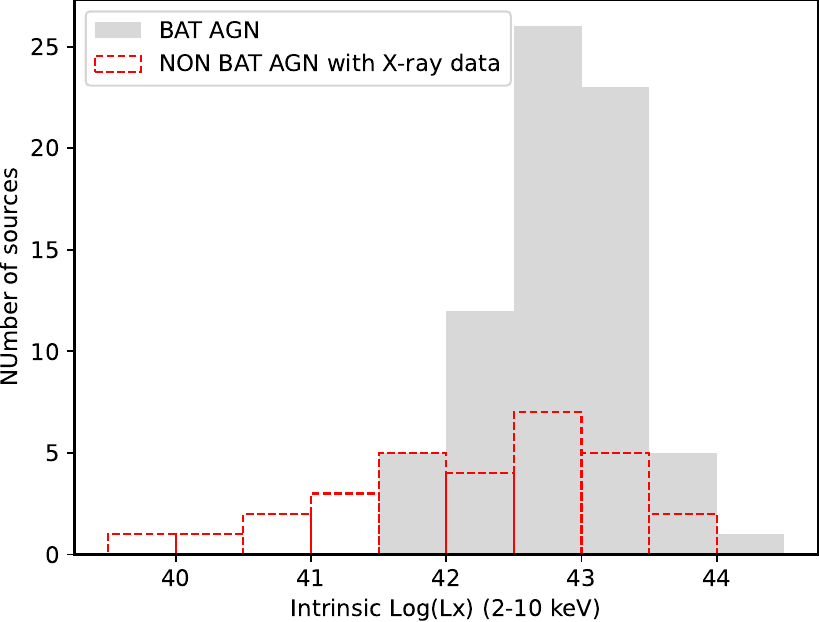}
\end{center}
\caption{Comparison of the 2-10 keV luminosity distribution for the sources detected in \swift/BAT 70 months all sky survey (grey-shaded histogram) with  those missed (red-dashed line).}  
\label{lx_dist}
\end{figure}

\begin{figure}
\begin{center}
\includegraphics[width=0.9\columnwidth]{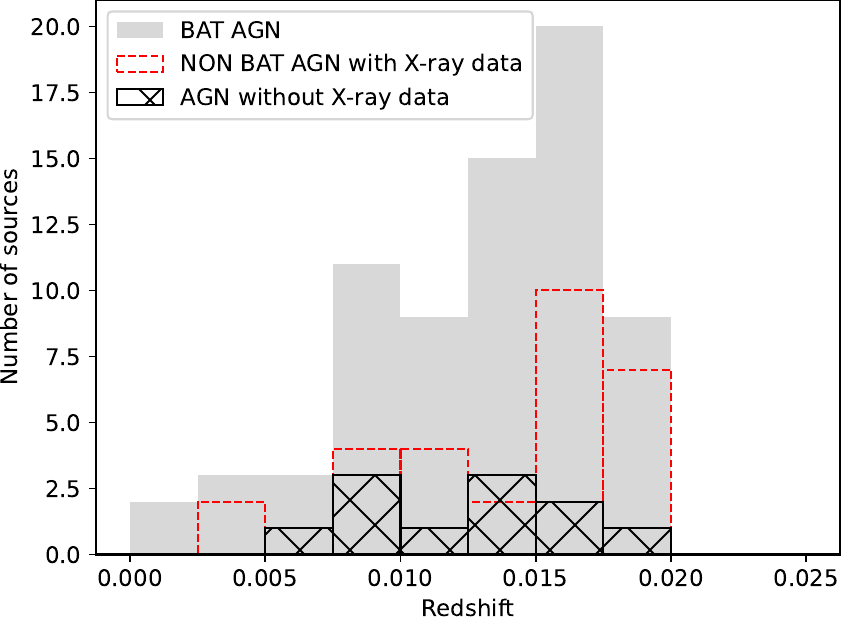}
\includegraphics[width=0.9\columnwidth]{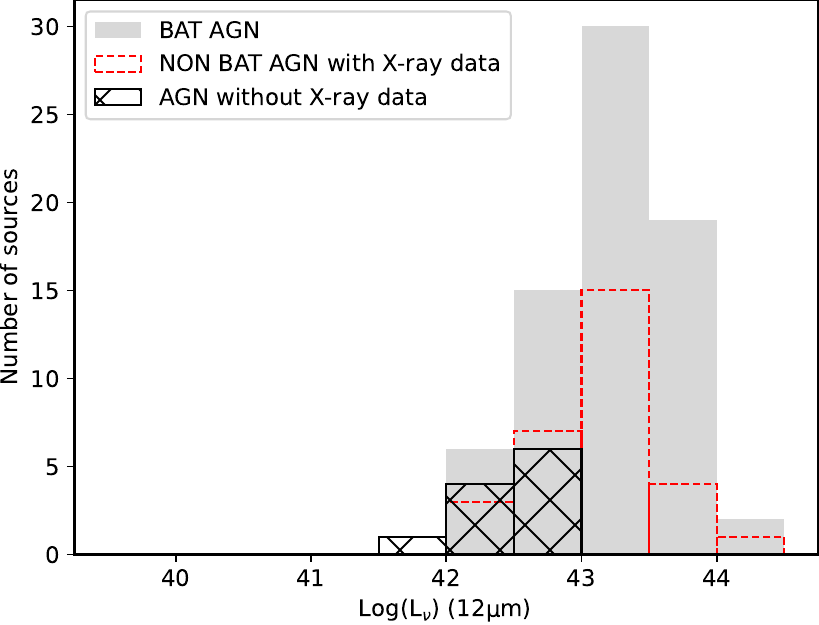}
\end{center}
\caption{(a) Upper panel: Comparison of the redshift distribution for the sources in our sample according to their origin. Sources found in BAT surveys are shown with gray histogram, sources with X-ray data missed by BAT survey are displayed with red-dashed line and sources without X-ray data are shown with the hatched histogram. (b) Lower panel: Similar as above for the $\rm L_{12 \mu m}$ luminosity obtained from \citet{asmus2020}.}  
\label{no_xrays}
\end{figure}

\section{Discussion}\label{discussion}
We organise our discussion as follows. In Section \ref{nhdistribution} we present the $N_H$ distribution 
 among the "known AGN" sample. In section \ref{noXraydata} we discuss about the prospects
 of finding a large number of Compton-thick AGN among the "new AGN" sample which has no BAT detections. It is likely that a number of Compton-thick  sources may lie among the  AGN that are not selected 
 by the {\it WISE} criteria. This is discussed in section \ref{belowAssefLine}. In section \ref{FaintLuminosities} we discuss the Compton-thick AGN that may be lurking among the fainter
  AGN sub-sample  with luminosities $\rm logL(12\mu m)[erg~s^{-1}]< 42.3$. In section \ref{DependenceLuminosity} we discuss the  
  possible dependence of the fraction of either 
  obscured or Compton-thick AGN on luminosity.
   Finally, in section \ref{ExtremeCT}
   we argue on the existence of extremely obscured 
   AGN with $\rm N_H>10^{25}~cm^{-2}$.

\subsection{The Compton-thick sources in the 'known AGN' sample}\label{nhdistribution}

While X-ray surveys offer the most unobscured view  towards the nucleus, when these sources are obscured by material close to or above the Compton thick limit - where the Compton scattering processes dominate over photo-ionization - the  X-ray sources suffer from significant attenuation hindering their detection. 
This results to a bias in the sense that  higher column density sources,  are under-represented in flux-limited surveys, even those in the BAT ultra hard X-ray band. Hence, the recovery of the true fraction of Compton-thick AGN, even in the local universe, remains problematic. 
Our analysis enables us to study the distribution of obscuration in the local universe in a nearly unbiased way as it is based on a complete volume limited galaxy sample. 

In Fig. \ref{nh_dist} we present the $\rm N_H$ distribution, along the line of site, for the sources in our sample. The upper panel compares the $\rm N_H$ distribution of the 70-month BAT AGN existing in our sample (grey-shaded histogram) with the $\rm N_H$ distribution of the sources missed by the hard X-ray survey (red-dashed line). The bottom panel shows the total $\rm N_H$ distribution of our sources.  
Notably, the vast majority of the sources missed by the BAT survey (14 out of 30) are associated with Compton-thick sources corresponding to a fraction of 45\%. This clearly suggests that 
below the BAT flux limit, lurks a population of Compton-thick AGN 
which evade detection. For all the 102 sources in our sample the fraction  of Compton-thick AGN  is 25/102 or 25$\pm$5\%.

In  Fig. \ref{lx_dist}   we compare the intrinsic 2-10 keV luminosity distribution 
of the BAT detected and non-detected sample. It is noteworthy  that a non-negligible number of sources within the non-BAT sample, extends to lower luminosities, $\rm L_x<10^{41.5} ~erg~s^{-1}$ forming a tail in the distribution. The vast majority of these sources are not
Compton-thick. Two of these have been analysed for the first time here, namely NGC3094 an UGC04145 showing no evidence for significant obscuration. Another source, (NGC2623) is moderately obscured \citep{yamada2021} while only 
NGC660 shows evidence for significant obscuration \cite{annuar2020}.
This suggests that a number of sources remain undetected by BAT because they have a  low intrinsic luminosity  
 rather than being associated with Compton-thick sources. Similar conclusions are drawn by \cite{yamada2021, Yamada2023} in a sample of nearby ultra-luminous infrared galaxies.

An unknown factor that may affect our estimations is the number of sources lacking X-ray information. For these sources the column density and the X-ray luminosity remain unknown, and as a result, they have not been included in the analysis. In Fig. \ref{no_xrays} we compare the redshift and the 12 micron luminosity, $\rm L_{12 \mu m}$, derived from \citet{asmus2020}, with the rest of our sample. The redshift distribution of the X-ray undetected sources shows no difference compared to the rest of the sample.  However, the $L_{12 \mu m}$ distribution clearly occupies the low luminosity part of 
the distribution $\rm L_{12 \mu m} < 10^{43} erg~s^{-1}$. This suggests that a number of sources have not been detected because they have low luminosities rather than they are associated with Compton-thick nuclei. 
 In the extreme case, where we assume that all eleven sources without X-ray information are associated with Compton-thick AGN, the fraction  would rise to 32$\pm$5\%.

\subsection{WISE selected AGN with no X-ray data available}\label{noXraydata}

\begin{figure}
\begin{center}
\includegraphics[width=0.9\columnwidth]{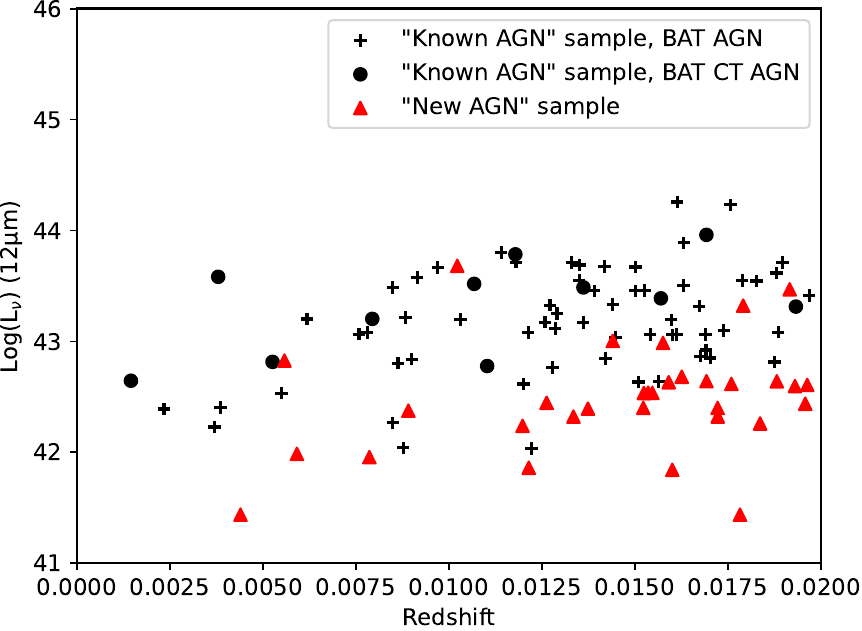}
\end{center}
\caption{The $\rm L_{12\mu m}$ vs redshift distribution of the BAT AGN against the "new" AGN sample.}
\label{comparizon_t2_t3}
\end{figure}

So far we have discussed the properties of the "known AGN" sample, containing as explained at section \ref{thesample},  {\it WISE} selected AGN, already known to host an active nucleus in the literature. As we mentioned \cite{asmus2020} define an additional AGN sample  or the "new AGN" sample, containing  {\it WISE} selected AGN, using the same criteria, but without any prior AGN classification in the literature. This sample comprises of  32 "new" AGN. 
  At first glance one would expect that the fraction of Compton-thick AGN is high and hence the 
  "new" AGN remain undetected by BAT. Indeed, the fraction of Compton-thick AGN in the "known AGN" sample that were undetected by BAT was particularly high of the order of 45\%. 
  The comparison  of the $\rm L_{12\mu m}$-z distribution of the BAT AGN and the "new AGN" sample  (Fig. \ref{comparizon_t2_t3})
     sheds more light on the reasons why the latter  sample may remain undetected in BAT. The distribution of the "new AGN" sample is  clearly skewed towards lower luminosities and higher redshift. 
     This suggests that a number of the "new" AGN 
     may remain undetected by BAT because of their low luminosity, provided that there is 
     a strong correlation between the $\rm L_{12\mu m}$ and the X-ray luminosity as found 
     in \cite{asmus2015} and \cite{stern2015}.  According to the above diagram, the most 
      strong candidates for being Compton-thick sources are a handful of sources 
       that mingle with the "known AGN"` sample in the same luminosity-redshift area.

\subsection{Compton-thick AGN missed by the WISE criteria}\label{belowAssefLine}

A number of well-known, BAT detected, Compton-thick AGN are not present in our sample. This is because of  the applied AGN selection criteria, based on {\it WISE} colours presented in section \ref{thesample}. 
For example a number of low luminosity AGN or those where the emission from the host galaxy dominates over the AGN may be missed by the W1-W2 criterion \citep[e.g.][]{Pouliasis2020}. Then, a question that  arises is whether our sample selection criteria affect the {\it fraction} of detected Compton-thick sources. 

In order to investigate this issue we examine the sample of Compton-thick AGN compiled by the Clemson group \citep[e.g.][]{Marchesi2018, zhao2021,torres2021}. These Compton-thick AGN were originally selected from the BAT AGN catalogue. Then the column density was accurately determined by means of \nustar observations. There are 
17 bona fide Compton-thick AGN in the Clemson sample\footnote{https://science.clemson.edu/ctagn/} within $z<0.02$. Only eight out of these 17 Compton-thick sources follow our selection criteria presented in section \ref{thesample}, while nine Compton-thick sources in the Clemson sample do not and therefore have been missed from this study. However, in a similar manner, in \swift/BAT 70 months all sky survey there are approximately 145 sources identified as AGN within z<0.02 according to \citet{baumgartner2013}. From this sample, half of the sources (72) satisfy the same {\it WISE} selection criteria while the other half (73) have been missed. This exercise clearly demonstrates that our selection criteria equally affect all the BAT population and not preferentially the highly obscured sources and therefore the measured fraction of Compton thick sources remains unaffected, at least for the fluxes probed by the \swift/BAT 70 months all sky survey. 

\subsection{Compton-thick AGN at faint luminosities}
\label{FaintLuminosities}
Up to this point, we have discussed the presence of Compton-thick sources
among luminous AGN with luminosities $\rm logL(12\mu m)[erg~s^{-1}] > 42.3$ regardless 
of their blue or red W1-W2 colours. 
However, \cite{asmus2020} point out that 70\% of known Seyferts have luminosities 
below this threshold. This means that an appreciable number of low-luminosity Compton-thick AGN may 
remain undetected. We use the relation of  hard X-ray luminosities vs. the nuclear MIR luminosities \cite[e.g.][]{Gandhi2009, Asmus2011} to convert our $\rm 12\mu m$ luminosity threshold to a 2-10 keV X-ray luminosity threshold. 
 We find that the corresponding 2-10 keV
 X-ray luminosity is approximately $\rm logL_X[erg~s^{-1}]>42.2$. 
 It has been suggested that the fraction of obscured AGN increases with decreasing luminosity or
 decreasing Eddington ratio
  \citep[e.g.][]{akylas2006, Ezhikode2017, ueda2014}. However, at very low luminosities this trend
  may be reversed. According to theoretical models no torus is formed 
  at very low bolometric luminosities, $\rm L_{BOL}<10^{42}~
  erg~s^{-1}$ \citep{Elitzur2006}.
  In any case, such a population of low luminosity Compton-thick AGN 
 may indeed exist without violating the hard X-ray and mid-IR background constraints \citep{comastri2015, nardini2011}. 
 We note here that,  \cite{boorman2024} presented simulations, suggesting that the new  High-Energy X-ray Probe-class mission concept 
 (HEX-P)  will be able to measure intrinsic luminosities and line-of-sight column densities but also distinguish between obscuration
geometries of this low luminosity population.

\subsection{Dependence of obscuration on luminosity}
\label{DependenceLuminosity}
\begin{figure}
\begin{center}
\includegraphics[width=0.9\columnwidth]{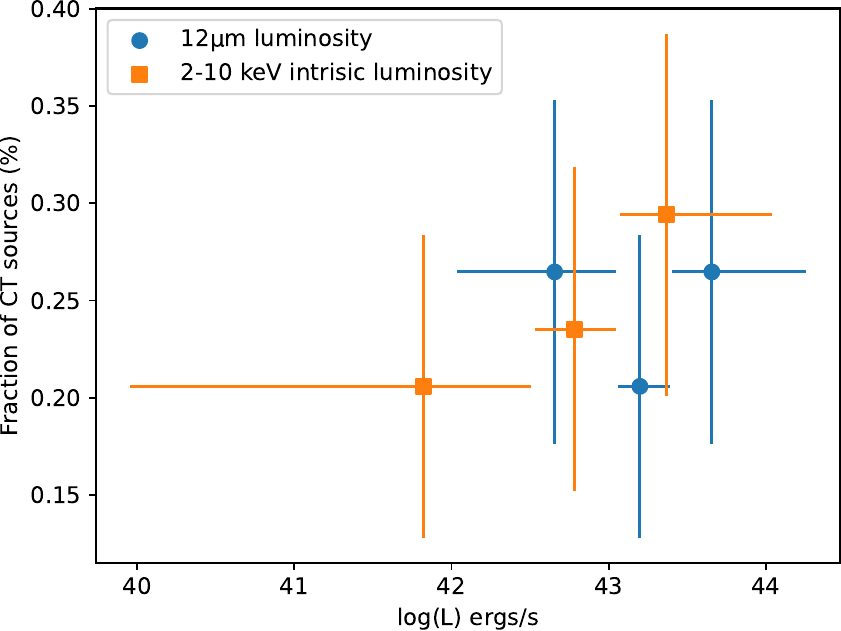}
\end{center}
\caption{Fraction of the Compton-thick sources ($\rm N_H>10^{24}~cm^{-2}$) in our sample as a function of the intrinsic 2-10 keV luminosity (blue-squared points) and the 12 micron luminosity (orange-circled points). The error in the estimated fraction corresponds to the 68 per cent confidence level while the uncertainty in the  luminosity axis denotes the range  of each luminosity bin used.}
\label{ct_vs_l}
\end{figure}

 Next, we investigate whether there is a dependence of the Compton-thick fraction on intrinsic 
 luminosity. If for example the fraction of Compton-thick AGN increases  with 
 decreasing luminosity, this could imply that the X-ray undetected sources in our sample harbour more heavily obscured sources.  Previous results on this subject remain controversial. In particular \citet{brightman2015}  found possible evidence of a strong decrease of the covering factor of the torus while \citet{buchner2015} found instead that the fraction of Compton-thick AGN is compatible with being constant with the X-ray luminosity. 
Also, \citet{ricci2015} analyzed the data from the \swift/BAT all-sky survey and provided the corrected for selection bias, intrinsic
column density distribution of Compton-thick AGN in the local universe in two different luminosity ranges. Their average estimated  fraction of Compton-thick sources is 27$\pm$4\%. They also present tentative evidence for a small decrease in the fraction of obscured Compton-thick AGN with increasing luminosity. 
In Fig. \ref{ct_vs_l}  we present the fraction of the Compton-thick sources  in our sample as a function of the 2-10 keV and the 12 micron luminosity.  The errors in the estimated 
Compton-thick fraction correspond to the 1$\sigma$ confidence level and the uncertainty in the  luminosity denotes the range  of each luminosity bin. Clearly, the plots show no evidence for any dependence of the Compton-thick fraction with luminosity.
Our results indicate a very similar fraction of Compton-thick sources at all luminosity bins, fully consistent with the average value of 25$\pm$5\%. However the limited statistics do not allow to rule out changes at the level of the quoted errors. 

\begin{figure}
\begin{center}
\includegraphics[width=0.9\columnwidth]{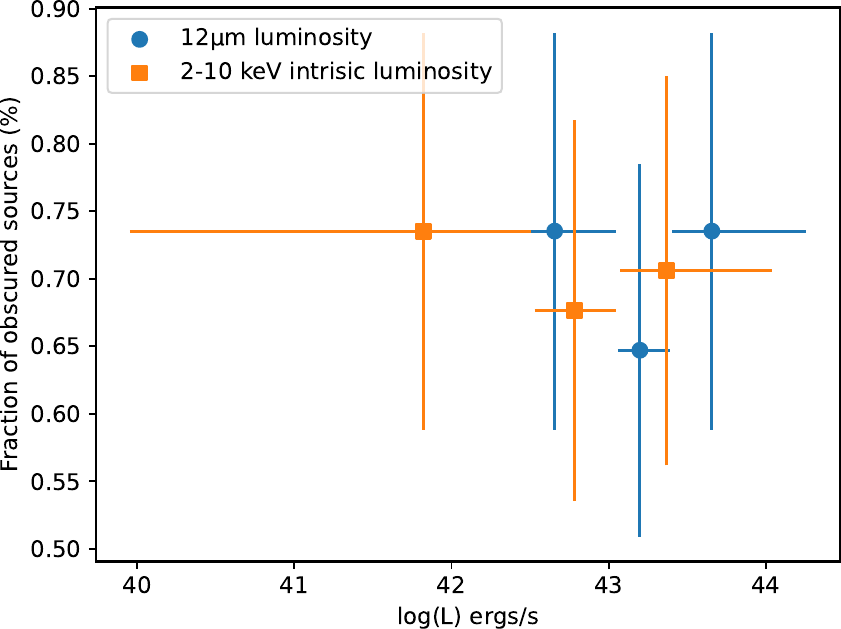}
\end{center}
\caption{Fraction of the obscured sources ($\rm N_H>10^{22}~cm^{-2}$)  as a function of the intrinsic 2-10 keV luminosity (blue-square points) and the 12 micron luminosity (orange-circle points). The error in the estimated fraction corresponds to the 68 per cent confidence level while the uncertainty in the  luminosity axis denotes the range  of each luminosity bin used.}
\label{obscured_vs_l}
\end{figure}

Next, we plot the  fraction of the obscured sources ($\rm N_H>10^{22}~cm^{-2}$) as a function of the intrinsic luminosity in Fig. \ref{obscured_vs_l}. Previous studies \cite[e.g.][]{ueda2003, akylas2006, buchner2015,ricci2015} suggest a clear decline in the fraction of obscuration with decreasing X-ray luminosity. Our analysis does not reveal any such trend in contradiction to the studies above. However, this trend has been previously identified across a much broader luminosity range than the one explored here. It is the highest luminosity bins that are not covered here, that play a  significant role in this trend \cite[e.g.][]{ueda2003}. 

 Alternatively, \citet{sazonov2015}, claim that this effect could be purely artificial due to a  negative bias in finding obscured sources and  a positive bias in finding  unobscured AGN, due to the reflected emission. According to these authors, the above biases lead to a decreasing observed fraction of obscured AGN with increasing luminosity even if there is no intrinsic luminosity dependence. In this scenario, the current analysis correctly finds a constant fraction of obscured sources in all 
 luminosities.

\subsection{Extreme obscuration  in the local universe} \label{ExtremeCT}

Our work so far has provide a robust and almost unbiased constraint of the Compton-thick fraction among the bright, WISE selected AGN in the local Universe. One key element, which has not been addressed, is the distribution of the column density 
among the Compton-thick sources.  Most X-ray background synthesis models  assume either a flat fraction of Compton-thick AGN over the entire range of $\rm log(N_H) [cm^{-2}]= 24-26 $  \citep[e.g.][]{ananna2019, ueda2014, gilli2007} or alternatively, all Compton-thick AGN are placed in the range of $\rm log(N_H) [cm^{-2}]= 24-25$ \citep[e.g.][]{akylas2016}.
Our observational leverage on the distribution remains uncertain. The only two kwown Compton-thick AGN  
which may have column densities close to  $\rm \sim N_H=10^{25} ~cm^{-2}$ are Circinus and NGC1068.
 The dearth of such highly  obscured sources 
 in the BAT surveys is possibly because the extreme obscuration prohibits their detection 
 even at distances as low as 100 Mpc \citep[e.g.][]{burlon2011}. 
 The limited $\rm N_H$ range of the current Compton-thick  spectral models, which typically have a ceiling in the maximum allowed $\rm N_H$ value of $\rm 10^{25}~cm^{-2}$, further complicates
 the secure identification of the most heavily obscured of the Compton-thick sources.

Our sample which does not suffer from any flux-limit bias offers the opportunity to further address this issue. 
We arbitrarily assume that all the Compton-thick sources with lower limit column density estimations $\rm N_H>10^{24}~cm^{-2}$, occupy the   $\rm log(N_H) [cm^{-2}]= 25-26 $ bin. 
All the other Compton-thick sources - those with a secure measurement in their $\rm N_H$ - are then  placed in the $\rm log(N_H) [cm^{-2}]= 24-25  $ bin. 
According  to  Table \ref{full_sample}, there are 25 Compton-thick sources in our sample and only in eight cases the $\rm N_H$ estimation is a lower limit. Then, this crude approximation shows that the fraction of $\rm log(N_H) [cm^{-2}]= 25-26$  sources  account roughly for at most  30\% of the Compton-thick population at least in the 'known AGN' sample.

\begin{table*}
\caption{Log of sources analysed in this work} 
\label{thenewsample}      
\centering                          
\begin{tabular}{c c c c c c c }   
\hline               
Source name  &  \multicolumn{2}{c}{\chandra} &   \multicolumn{2}{c}{\xmm}  & \multicolumn{2}{c}{\nustar} \\ 
             &     OBSID  & Exp. (s) & OBSID  & Exp. (s) &  OBSID  & Exp. (s) \\
\hline               
2MASXJ01500266-0725482 &    - & -     &0200431101 & 11921 & 60360005002 & 30725 \\
2MASXJ04405494-0822221 &    - &  -    &0890690401 & 18000 & 60701043002 & 30706 \\
2MASXJ04524451-0312571 &    - &  -    &0307002501 & 18111 &     -       &  - \\
CGCG074-129            &    - &  -    &0822391201 & 13600 &     -       &  - \\
ESO018-G009            &    - &  -    &0805150401 & 19200 & 60362029002 & 28755  \\
ESO420-013$^{1}$  &10393 & 12760 & -          & - & 60668003002 & 108164 \\
IC4769                 &    - &  -    &0405380501 & 35013 &    -        & \\
IC4995$^{2}$       &    - &  -    &0200430601 & 11912 & 60360003002 & 33998 \\
NGC3094                &    - &  -    &0655380801 & 16918 & 60668001002 & 100823 \\
NGC5990                &    - &  -    &0655380901 & 18918 &   -         & \\
UGC01214               &    - &  -    &0200430701 & 11913 & 60360004002 & 31998 \\
UGC04145               &    - &  -    &0763460201 & 18000 &   -         & -     \\

\hline  
\multicolumn{5}{l}{\small $^1$ For this source  \xmm  observations where not available. Instead we make use of \chandra data.} \\
\multicolumn{5}{l}{\small $^2$ X-ray spectral analysis has been presented in \cite{clavijo2022}. Here we have repeated } \\
\multicolumn{5}{l}{\small  the analysis using spectral modeling suitable to the Compton thick nature of the source.} \\
\end{tabular}
\end{table*}

\section{Summary}
We analysed the X-ray properties of the \cite{asmus2020} paper of local ($z<0.02$) AGN. Our basic goal is to constrain the number  density of Compton-thick sources. We primarily focus
on the AGN sample selected on the basis of the {\it WISE} W1 and W2 colours. This is divided in two subsamples. The “known AGN” sample,  already known to host an active nucleus in the literature, that contains 113 sources of which the vast majority (102) have been observed by various X-ray missions  (72 have been detected by BAT) and the “new AGN” sample which contains 32 sources which have no prior AGN classification in the literature. For the first sample, we compile the X-ray observations available in the literature. We also analyse here for the first time, the  \nustar, \xmm and \chandra observations of eleven sources. As the sample examined here is not flux-limited, it provides us with the best opportunity so far to study the full Compton-thick AGN population.
Our results can be summarised as follows.

\begin{itemize}
\item{Our spectral analysis employing both the {\sc RXTORUS} and the {\sc UXCLUMPY} models reveals four new Compton-thick sources with column densities in excess of $\rm 4\times 10^{24} cm^{-2}$} 
\\
\item{The fraction of Compton-thick sources among the 102 sources with available X-ray data 
in the "known AGN" sample is 25$\pm$5\%. Even in the extreme case where all the sources with no available X-ray data were associated with Compton-thick AGN the Compton-thick fraction would rise 
 to 31$\pm$5 \%} \\
\item{The fraction of Compton-thick AGN among the 30 sources that have not been detected by BAT is
  much higher (44\%) compared to the fraction of the Compton-thick sources in the BAT detected sources which is only 16\%.} \\
  \item{Regarding the "new AGN" sample, we argue that most of these sources have not been detected by BAT because they have low luminosity rather than high obscuration.} \\
\end{itemize}

\begin{landscape}
\begin{table}
\caption{Spectral fitting results using the simple power-law model} 
\label{results_simple_fit}      
\centering                          
\begin{tabular}{c c c c c c c c c c  c}   
\hline               
Source name & Mission$^1$ & \multicolumn{9}{c}{$\rm PHABS_{GAL}*(APEC+PHABS*PO+GA)$ model parameters}  \\   
\hline               
 & & kT & $\rm N_{APEC}$ & $\rm \Gamma_{soft~PO}$ & $\rm N_{soft~PO}$ & $\rm \Gamma_{hard~PO}$  & $N_H$ & $\rm N_{hard~PO}$ & $\rm FeK~line~(EW)$ & $\rm \chi^2$/dof \\    
 & &keV & $\times 10^{-5}$ &  &  $\times 10^{-5}$ &  & $\times 10^{22} cm^{-2}$ & $\times 10^{-3}$ & eV &     \\    
\hline                
 2MASXJ01500266-0725482 & X,N & - & - & - & - & $2.11^{+0.10}_{-0.10}$ & $0.091^{+0.34}_{-0.33}$ & $0.36^{+0.08}_{-0.06}$ & $6.43^{+0.25}_{-0.10}$ ($290^{+190}_{-150})$ &  115.99/103  \\
 
 
 NGC3094 & X,N & $0.77^{+0.10}_{-0.12}$ & $0.87^{+0.74}_{-0.35}$ & - & - &  $1.92^{+0.41}_{-0.28}$ & <0.14 & $0.021^{+0.018}_{-0.011}$ & - &  48.89/40  \\
 
 UGC04145 & X & $0.68^{+0.11}_{-0.13}$ & $1.24^{-0.69}_{-0.64}$ & - & - &  $2.43^{+1.55}_{-0.83}$ & <1.16 & $0.019^{+0.073}_{-0.010}$ & - &  6.7/8  \\ 
 
\hline       
\multicolumn{11}{l}{\small $^1$ X:\xmm data have been used , C: \chandra data have been used, N: \nustar data have been used} \\
\end{tabular}
\end{table}

\begin{table}
\caption{Spectral fitting results using the {\sc RXTORUS} model} 
\label{results_advanced_model}      
\centering                          
\begin{tabular}{c c c c c c c c c c c c c}   
\hline               
Source name  & Mission$^1$ & \multicolumn{11}{c}{$\rm PHABS_{GAL}*(PO+APEC+RXTORUS)$ model parameters}  \\   
\hline               
    && $\rm \Gamma_{soft}$ & $\rm N_{soft}$ & kT & $\rm N_{APEC}$ & $\rm \Gamma$  & $\rm N_H^{eq}$ & $\rm N_H^{los}$ & $\rm \theta_{inc}$ & $\rm sin(\phi_{torus})$ &  $\rm N_{RXTORUS}$ & $\rm \chi^2$/dof \\    
    &&   & $\times 10^{-5}$ &    keV           & $\times 10^{-5}$      &               & $\rm \times 10^{22} cm^{-2}$ & $\rm \times 10^{22} cm^{-2}$ & deg. &   & $\times 10^{-3}$ &      \\    
\hline                

ESO420-013 & C,N & $3.77^{+0.37}_{-0.29}$ & $5.83^{+2.79}_{-2.51}$ & $0.89^{+0.08}_{-0.09}$   & $3.14^{+2.20}_{-1.23}$  & 2(f) & >508 & >300 & $77^{-6}_{-2}$ & 0.5(f) & $9.84^{+6.75}_{-7.39}$ & 101.2/96 \\

ESO018-G009 & X,N & - & - & $0.75^{+0.10}_{-0.10}$   & $0.91^{+0.19}_{-0.12}$  & $2(f)$ & $31^{+24}_{-17}$ & $31^{+24}_{-17}$ & 90(f) & 0.5(f) & $10.77^{+2.30}_{-4.41}$ & 40.51/31 \\

UGC01214 & X,N & $4.52^{+0.42}_{-0.33}$ & $14.34^{+4.91}_{-3.86}$ & $0.92^{+0.09}_{-0.08}$   & $4.78^{+1.92}_{-1.63}$  & $2(f)$ & >598 & >505  & 69$^{+4}_{-3}$ & 0.5(f) & $7.01^{+2.62}_{-1.1}$ & 89.59/73 \\

2MASXJ04405494-0822221 & X,N & - & - & $0.33^{+0.22}_{-0.07}$  & 0.54$^{+0.25}_{-0.23}$  & 2(f) & >421 & >300 & 69$^{+6}_{-10}$ & 0.5(f) & 4.16$^{+8.42}_{-1.33}$ & 74.82/58 \\

 CGCG074-129  & X         & 2(f) & 0.67$^{+0.23}_{-0.24}$ & -  & - & 2(f) & 68$^{+84}_{-37}$ & 68$^{+84}_{-37}$ & 90(f) & 0.5(f) & 0.58$^{+0.43}_{-0.43}$ & 23.01/21 \\

 2MASXJ04524451-0312571 & X &2(f) & 1.24$^{+0.26}_{-0.26}$ &  1.01$^{+0.34}_{-0.28}$ & 0.22$^{+0.23}_{-0.19}$ & 2(f) & >40 & >40 & 90(f) & 0.5(f) & 0.27$^{+4.61}_{-0.25}$ & 12.23/14 \\

 IC4769   & X              & 2(f) & 1.16$^{+0.24}_{-0.24}$ &  0.76$^{+0.10}_{-0.10}$ & 0.47$^{+0.14}_{-0.14}$ & 2(f) &  362$^{+577}_{-116}$  & 362$^{+577}_{-116}$ & 90(f) & 0.5(f) & 35.32$^{+210.57}_{-22.96}$ & 45.89/49 \\

 IC4995 & X,N                & 6.01$^{+2.41}_{-1.32}$   & 0.94$^{+1.15}_{-0.71}$ &  0.82$^{+0.13}_{-0.13}$ & 1.04$^{+0.59}_{-0.55}$ & 2.07$^{+0.16}_{-0.11}$ &  >815  & >482 & $64^{+13}_{-6}$ & $0.5(f)$ & 9.53$^{+14.19}_{-2.80}$ & 66.62/54 \\

 NGC5990 & X              & 2.91 & 2.40$^{+0.42}_{-0.42}$ &  0.75$^{+0.11}_{-0.11}$ & 1.21$^{+0.29}_{-0.29}$ & 2(f) &  27$^{+26}_{-12}$  & 27$^{+26}_{-12}$  & 90(f) & 0.5(f) & 0.12$^{+0.14}_{-0.05}$ & 30.36/30 \\

\hline      
\multicolumn{13}{l}{\small $^1$ X:\xmm data have been used , C: \chandra data have been used, N: \nustar data have been used} \\

\end{tabular}
\end{table}

\begin{table}
\caption{Spectral fitting results of Compton-thick sources only using the {\sc UXCLUMPY} model} 
\label{results_uxclumpy_model}      
\centering                          
\begin{tabular}{c c c c c c c c c c c c c}   
\hline               
Source name  & Mission$^1$ & \multicolumn{11}{c}{$\rm PHABS_{GAL}*(PO+APEC+UXCLUMPY)$ model parameters}  \\   
\hline               
    && $\rm \Gamma_{soft}$ & $\rm N_{soft}$ & kT & $\rm N_{APEC}$ & $\rm \Gamma$  & $\rm N_H^{eq}$ & $\rm N_H^{los}$ & $\rm \theta_{inc}$ & $\rm (\sigma_{torus})$ &  $\rm N_{UXCLUMPY}$ & $\rm \chi^2$/dof \\    
    &&   & $\times 10^{-5}$ &    keV           & $\times 10^{-5}$      &               & $\rm \times 10^{22} cm^{-2}$ & $\rm \times 10^{22} cm^{-2}$ & deg. &   & $\times 10^{-3}$ &      \\    
\hline                

ESO420-013 & C,N & $3.35^{+0.40}_{-0.24}$ & $5.14^{+2.06}_{-2.43}$ & $0.86^{+0.08}_{-0.07}$   & $2.55^{+0.44}_{-1.21}$  & $1.53^{+0.27}_{-0.28}$ & - & >292 & 90(f) & 30(f) & $0.89^{+1.42}_{-0.11}$ & 100.83/95 \\

UGC01214 & X,N & $3.61^{+0.20}_{-0.16}$ & $9.61^{+0.91}_{-1.01}$ & $0.88^{+0.10}_{-0.09}$   & $1.58^{+0.60}_{-0.58}$  & $2.33^{+0.31}_{-0.22}$  & - & >345  & 90(f) & 30(f) & $16.46^{+21.70}_{-7.91}$ & 85.17/72 \\

2MASXJ04405494-0822221 & X,N & - & - & $0.38^{+0.33}_{-0.12}$  & 0.22$^{+0.18}_{-0.14}$  & $2.17^{+0.20}_{-0.21}$ & - & >298 & 90(f) & 30(f) & 18.57$^{+11.21}_{-6.90}$ & 80.91/58 \\

IC4769   & X              & - & - &  0.75$^{+0.10}_{-0.10}$ & 0.38$^{+0.11}_{-0.11}$ & $2.01^{+0.34}_{-0.32}$ &  -  & >142 & 90(f) & 30(f) & 5.12$^{+6.47}_{-2.97}$ & 71.42/55 \\

IC4995 & X,N              & 5.53$^{+3.15}_{-1.62}$   & 0.81$^{+1.41}_{-0.69}$ &  0.79$^{+0.16}_{-0.14}$ & 0.73$^{+0.44}_{-0.42}$ & 2.38$^{+0.27}_{-0.11}$ &  -  & >363 &   90(f) & 30(f) & $36.6^{+28.17}_{-11.53}$ & 86.72/53 \\

\hline      
\multicolumn{13}{l}{\small $^1$ X:\xmm data have been used , C: \chandra data have been used, N: \nustar data have been used} \\

\end{tabular}
\end{table}

\end{landscape}


\begin{table*}
\caption{Flux and Luminosity of the sources fitted with the simple model} 
\label{fluxes_simple_fit}      
\centering                          
\begin{tabular}{l c c c c c c c c }   
\hline    
Source name  & \multicolumn{4}{c}{Intrinsic flux $\rm \times 10^{-13} ergs ~s^{-1} ~cm^{-2}$} & \multicolumn{4}{c}{Intrinsic luminosity $\rm \times 10^{42} ~ergs~s^{-1}$} \\ \\ 
\hline 
             & $\rm F^{APEC}_{0.5-2~keV}$ & $\rm F^{PL}_{0.5-2~keV}$ & $\rm F^{PL}_{2-10~keV}$ & $\rm F^{PL}_{10-80~keV}$ & $\rm L^{APEC}_{0.5-2~keV}$ & $\rm L^{PL}_{0.5-2~keV}$ & $\rm L^{PL}_{2-10~keV}$ & $\rm L^{PL}_{10-80~keV}$ \\ \\   
\hline  
 2MASXJ01500266-0725482 &   -  &  8.38 & 7.90 & 2.01  &   -   & 0.585   & 0.554  & 0.145  \\
 NGC3094                & 0.25 &  0.48 & 0.63 & 0.94  & 0.004 & 0.007   & 0.009  & 0.013 \\
 UGC04145               & 0.37 &  0.44 & 0.26 &  -    &  0.019& 0.022   & 0.014  & -    \\

\hline                                                  
\end{tabular}
\end{table*}

\begin{table*}
\caption{Flux and Luminosity: {\sc RXTORUS}  model.} 
\label{fluxes_advanced_fit}      
\centering                          
\begin{tabular}{l c c c c c c c c }   
\hline    
Source name  & \multicolumn{4}{c}{Intrinsic flux $\rm \times 10^{-13} ergs ~s^{-1} ~cm^{-2}$} & \multicolumn{4}{c}{Intrinsic luminosity $\rm \times 10^{42} ~ergs~s^{-1}$} \\ \\ 
\hline 
             & $\rm F^{APEC}_{0.5-2~keV}$ & $\rm F^{PL, soft}_{0.5-2~keV}$ & $\rm F^{RXTORUS}_{2-10~keV}$ & $\rm F^{RXTORUS}_{10-80~keV}$ & $\rm L^{APEC}_{0.5-2~keV}$ & $\rm L^{PL, soft}_{0.5-2~keV}$ & $\rm L^{RXTORUS}_{2-10~keV}$ & $\rm L^{RXTORUS}_{10-80~keV}$ \\ \\   
\hline  
ESO420-013              & 0.83 & 1.64  & 255.75  &  17.28  & 3.050  & 7.51  &  25.08  & 1.88  \\
ESO018-G009             & 2.62 &   -   & 2.71    &   0.76  & 0.017 &   -    &  0.176  & 0.051 \\
UGC01214                & 0.62 & 3.16 & 123.61  &  34.7   & 0.020 &  0.102  &   8.04  & 12.09  \\
2MASXJ04405494-0822221  & 0.10 &   -   & 150.31  &   53.88 & 0.005 &   -    &   7.56  & 13.12 \\
CGCG074-129             & -    & 11.71 &  14.14  &      -  &   -   &  0.009 &   0.81  &  -    \\
2MASXJ04524451-0312571  & 0.05 & 0.27  &   6.79  &      -  &  0.003 & 0.015 &   0.37  &  -    \\
IC4769                  & 0.14 & 0.25  & 888.55  &      -  &  0.007 & 0.013 &  44.86  &  -    \\
IC4995                  & 0.29 & 0.60  & 214.89  &   56.02 &  0.017 & 0.039 &  13.17  &  3.55 \\
NGC5990                 & 0.35 & 0.57  &  2.96   &      -  &  0.011  & 0.019 &  0.10  &  -    \\

\hline                                                  
\end{tabular}
\end{table*}

\begin{acknowledgements}
This research is based on observations obtained with \xmm, an ESA science mission with instruments and contributions directly funded by ESA Member States and NASA. This research has made use of data from the \nustar mission, a project led by the California Institute of Technology, managed by the Jet Propulsion Laboratory, and funded by the National Aeronautics and Space Administration. Data analysis was performed using the \nustar Data Analysis Software (NuSTARDAS), jointly developed by the ASI Science Data Center (SSDC, Italy) and the California Institute of Technology (USA). This research has made use of data obtained from the Chandra Data Archive and the Chandra Source Catalog, and software provided by the Chandra X-ray Center (CXC) in the application packages CIAO and Sherpa.  This research has made use of data and software provided by the High Energy Astrophysics Science Archive Research Center (HEASARC), which is a service of the Astrophysics Science Division at NASA/GSFC.
This research uses data supplied by the UK Swift Science Data Centre at the University of Leicester. 
\end{acknowledgements}

\bibliography{ref}
\bibliographystyle{aa}

\begin{appendix}
\onecolumn

\begin{landscape}
\section{The sample}

\begin{longtable}{lccccccccc}
\caption{\label{full_sample} Log of our sample} \\
\hline\hline
Name & class &  Ra & Dec &	z & logL(W3) & $\rm logL_{nuc}$ & $\rm L_{2-10~keV}$ & $\rm NH_{los}$ & Ref. \\
(I)  & (II) &  (III) & (IV) &	(V) & (VI) & (VII) & (VIII) & (IX) & (X) \\

\hline 
\endfirsthead
\caption{continued.}\\
\hline\hline
Name & class &  Ra & Dec &	z & logL(W3) & $\rm logL_{nuc}$ & $\rm L_{2-10~keV}$ & $\rm N_H^{los}$ & Ref. \\
(I)  & (II) &  (III) & (IV) &	(V) & (VI) & (VII) & (VIII) & (IX) & (X) \\

\hline
\endhead
\hline
\endfoot
2MASXJ00253292+6821442$^{\star}$ & Sy2 & 6.387 & 68.3623 & 0.012 & 42.86 & 42.62 & 43.16 & 23.98 & 1 \\
NGC0262$^{\star}$ & Sy2 & 12.1964 & 31.957 & 0.015 & 43.62 & 43.45 & 43.47 & 23.12 & 1 \\
NGC0424$^{\star}$ & Sy2 & 17.8651 & -38.0835 & 0.012 & 43.82 & 43.78 & 42.62 & 24.4 & 2 \\
NGC0454NED02$^{\star}$ & Sy2 & 18.6039 & -55.3971 & 0.012 & 43.04 & 43.08 & 42.2 & 23.3 & 1 \\
NGC0449 & Sy2  & 19.0302 & 33.0896 & 0.016 & 43.42 & 43.23 & 41.89 & 24.05 & 3 \\
NGC0526A$^{\star}$ & Sy2  & 20.9766 & -35.0655 & 0.019 & 43.5 & 43.71 & 43.29 & 22.01 & 1 \\
UGC01032$^{\star}$ & Sy1 & 21.8856 & 19.1788 & 0.017 & 43.29 & 43.1 & 42.73 & 20.61 & 1 \\
NGC0660 & Sy2  & 25.76 & 13.6451 & 0.003 & 42.84 & 42.43 & 41.00 & >23.00 & 4 \\
UGC01214 & Sy2  & 25.9908 & 2.3499 & 0.017 & 43.67 & 43.58 & 42.90 & >24.7 & This work \\ 
2MASXJ01500266-0725482 & Sy2  & 27.5112 & -7.4301 & 0.018 & 43.52 & 43.35 & 41.74 & 21.56 & This work \\
NGC0788$^{\star}$ & Sy2 & 30.2769 & -6.8155 & 0.014 & 43.21 & 43.17 & 43.05 & 23.82 & 1 \\
MRK1044$^{\star}$ & Sy1 & 37.523 & -8.9981 & 0.016 & 43.26 & 43.06 & 42.49 & 20 & 1 \\
MESSIER077$^{\star}$ & Sy2 & 40.6696 & -0.0133 & 0.004 & 44.26 & 43.58 & 43.34 & >25 & 5 \\
MRK1058 & Sy2  & 42.466 & 34.988 & 0.017 & 43.09 & 42.87 & - & - & - \\
NGC1125$^{\star}$ & Sy2 & 42.9185 & -16.6507 & 0.011 & 43.01 & 42.78 & 42.4 & 24.06 & 6 \\
MCG-02-08-014$^{\star}$ & Sy2 & 43.0975 & -8.5104 & 0.017 & 42.88 & 42.87 & 42.86 & 23.03 & 1 \\
UGC02456 & Sy2  & 44.9941 & 36.8206 & 0.012 & 43.55 & 43.38 & 42.08 & 23.98 & 7 \\
NGC1194$^{\star}$ & Sy2 & 45.9546 & -1.1037 & 0.014 & 43.4 & 43.49 & 42.78 & 24.15 & 8 \\
NGC1275$^{\star}$ & Sy2  & 49.9507 & 41.5117 & 0.018 & 44.16 & 44.23 & 43.99 & 21.68 & 1 \\
NGC1320 & Sy2  & 51.2029 & -3.0423 & 0.009 & 43.13 & 42.91 & 42.85 & 24.6 & 14 \\
ESO116-G018 & Sy2  & 51.221 & -60.7384 & 0.018 & 43.57 & 43.41 & 43.23 & 24.41 & 12 \\
NGC1365$^{\star}$ & Sy1 & 53.4016 & -36.1404 & 0.006 & 43.28 & 42.53 & 42.07 & 22.21 & 1 \\
NGC1386 & Sy1 & 54.1924 & -35.9994 & 0.003 & 42.55 & 42.37 & 41.90 & 24.7 & 8 \\
ESO548-G081$^{\star}$ & Sy1 & 55.5153 & -21.2443 & 0.014 & 43.21 & 43.04 & 43.03 & 20 & 1 \\
ESO420-G013 & Sy2  & 63.457 & -32.007 & 0.012 & 43.62 & 43.21 & 43.43 & >24.65 & This work \\
2MASXJ04405494-0822221 & Sy2  & 70.229 & -8.3728 & 0.015 & 43.76 & 43.61 & 42.88 & >24.47 & This work \\
UGC03157$^{\star}$ & Sy2 & 71.6236 & 18.4609 & 0.015 & 43.26 & 43.06 & 42.88 & 23.78 & 1 \\
2MASXJ04524451-0312571 & Sy2  & 73.1853 & -3.2159 & 0.016 & 43.28 & 43.08 & 41.57 & >23.6 & This work \\
ESO033-G002$^{\star}$ & Sy2  & 73.9957 & -75.5412 & 0.018 & 43.60 & 43.54 & 42.16 & 22.51 & 1 \\
GALEXASCJ051045.55+162958.9$^{\star}$ & Sy1 & 77.6896 & 16.4989 & 0.018 & 43.7 & 43.55 & 43.55 & 21.8 & 1 \\
2MFGC04298$^{\star}$ & Sy2  & 79.0947 & 19.4531 & 0.019 & 43.04 & 42.81 & 42.45 & 22.9 & 1 \\
ESO362-G018$^{\star}$ & Sy2 & 79.8992 & -32.6576 & 0.013 & 43.22 & 43.17 & 42.99 & 20 & 1 \\
NGC2110$^{\star}$ & Sy2 & 88.0474 & -7.4562 & 0.008 & 43.16 & 43.08 & 42.75 & 22.94 & 1 \\
UGC03426$^{\star}$ & Sy2 & 93.9015 & 71.0375 & 0.014 &  & 43.69 & 44.04 & 23.9 & 9 \\
UGC03478$^{\star}$ & Sy1 & 98.1965 & 63.6737 & 0.013 & 42.99 & 42.76 & 42.46 & 21.18 & 1 \\
NGC2273 & Sy2  & 102.5361 & 60.8458 & 0.006 & 43.01 & 42.78 & 43.11 & >24.85 & 8 \\
IC0450$^{\star}$ & Sy1 & 103.0511 & 74.4271 & 0.019 & 43.76 & 43.62 & 43.12 & 20.76 & 1 \\
UGC03752$^{\star}$ & Sy2 & 108.5161 & 35.2793 & 0.016 & 43.56 & 43.39 & 42.77 & 24.02 & 9 \\
2MASXJ07170726-3254197 & AGN(?) & 109.2802 & -32.9054 & 0.008 & 42.5 & 42.22 & - & - & - \\
CGCG147-020$^{\star}$ & Sy2 & 111.4057 & 29.9541 & 0.019 & 43.4 & 43.08 & 43.32 & 23.86 & 1 \\
UGC03995B$^{\star}$ & Sy2 & 116.038 & 29.2474 & 0.016 & 43.38 & 43.19 & 42.47 & 23.92 & 1 \\
IC2207$^{s}$ & Sy2  & 117.462 & 33.9623 & 0.016 & 42.83 & 42.41 & 42.23 & 20 & This work  \\
UGC04145 & Sy2  & 119.9172 & 15.3868 & 0.016 & 43.53 & 43.36 & 40.15 & 20 & This work  \\
PhoenixGalaxy$^{\star}$ & Sy2 & 121.0244 & 5.1139 & 0.013 & 43.7 & 43.55 & 43.16 & 23.4 & 1 \\
MCG-02-22-003$^{s}$ & Sy2  & 125.3898 & -13.3511 & 0.014 & 42.76 & 42.51 & 42.63 & 23.86 & This work \\ 
ESO018-G009 & Sy2  & 126.0329 & -77.7826 & 0.018 & 43.43 & 43.25 & 41.23 & 23.49 & This work \\
NGC2623 & LINER & 129.6003 & 25.7546 & 0.018 & 43.58 & 43.55 & 40.9 & 22.78 & 10 \\
ARP007 & Sy1 & 132.5843 & -16.5795 & 0.019 & 43.14 & 42.78 & - & - & - \\
MCG-01-24-012$^{\star}$ & Sy2  & 140.1927 & -8.0561 & 0.02 & 43.34 & 43.41 & 43.27 & 22.81 & 1 \\
ESO434-G040$^{\star}$ & Sy2 & 146.9173 & -30.9487 & 0.008 & 43.45 & 43.48 & 43.23 & 22.18 & 1 \\
MRK1239 & Sy1 & 148.0796 & -1.6121 & 0.02 & 44.15 & 44.12 & 42.93 & 23.54 & 11 \\
NGC3094 & AGN(?) & 150.3581 & 15.7701 & 0.008 & 43.5 & 43.66 & 39.95 & 20 & This work \\ 
CGCG064-055 & Sy2  & 151.4633 & 12.9613 & 0.009 & 42.74 & 42.48 & - & - & - \\
NGC3227$^{\star}$ & Sy1 & 155.8774 & 19.8651 & 0.004 & 42.83 & 42.41 & 42.28 & 20.95 & 1 \\
ESO317-G041$^{\star}$ & Sy2  & 157.8463 & -42.0606 & 0.019 & 43.6 & 43.31 & 42.5 & 24.31 & 6 \\
NGC3281$^{\star}$ & Sy2 & 157.967 & -34.8537 & 0.011 & 43.67 & 43.52 & 42.98 & 24.3 & 12 \\
NGC3516$^{\star}$ & Sy1 & 166.6979 & 72.5686 & 0.009 & 43.4 & 43.22 & 42.98 & 20 & 1 \\
IRAS11215-2806 & Sy2  & 171.0114 & -28.3877 & 0.014 & 43.21 & 43 & - & - & - \\
NGC3783$^{\star}$ & Sy1 & 174.7573 & -37.7387 & 0.01 & 43.67 & 43.67 & 43.55 & 20.49 & 1 \\
NGC4051$^{\star}$ & Sy1 & 180.7901 & 44.5313 & 0.002 & 42.41 & 42.39 & 41.59 & 20 & 1 \\
NGC4253$^{\star}$ & Sy1 & 184.6105 & 29.8129 & 0.013 & 43.5 & 43.33 & 42.73 & 20.32 & 1 \\
NGC4388$^{\star}$ & Sy2 & 186.4448 & 12.6621 & 0.008 & 42.66 & 42.26 & 42.45 & 23.52 & 1 \\
NGC4507$^{\star}$ & Sy2 & 188.9026 & -39.9093 & 0.012 & 43.67 & 43.71 & 43.54 & 23.95 & 1 \\
NGC4593$^{\star}$ & Sy1 & 189.9143 & -5.3443 & 0.009 & 42.93 & 42.84 & 42.64 & 20 & 1 \\
IC3639 & Sy2  & 190.2202 & -36.7559 & 0.011 & 43.52 & 43.44 & 43.4 & 24.56 & 13 \\
NGC4628 & Sy2  & 190.6053 & -6.971 & 0.009 & 43.12 & 42.9 & - & - & - \\
ESO323-G077$^{\star}$ & Sy2 & 196.6089 & -40.4147 & 0.015 & 43.86 & 43.67 & 42.91 & 22.81 & 1 \\
NGC4968 & Sy2  & 196.7749 & -23.677 & 0.01 & 43.24 & 43.03 & 42.88 & 24.3 & 15 \\
MCG-03-34-064$^{\star}$ & Sy2 & 200.6019 & -16.7285 & 0.017 & 44.13 & 43.96 & 43.46 & 24.13 & 20 \\
NGC5135 & Sy2  & 201.4336 & -29.8337 & 0.014 & 43.77 & 43.17 & 43.3 & 24.78 & 10 \\
ESO383-G018$^{\star}$ & Sy2 & 203.3588 & -34.0148 & 0.013 & 43.43 & 43.12 & 42.81 & 23.31 & 1 \\
ESO383-G035$^{\star}$ & Sy1 & 203.9738 & -34.2955 & 0.008 & 43.07 & 43.06 & 42.75 & 20.85 & 1 \\
2MASSJ13473599-6037037$^{\star}$ & Sy2 & 206.8998 & -60.6177 & 0.013 & 43.43 & 43.25 & 43.28 & 22.9 & 1 \\
IC4329A$^{\star}$ & Sy1 & 207.3303 & -30.3094 & 0.016 & 44.21 & 44.26 & 43.87 & 21.52 & 1 \\
2MASXJ13512953-1813468$^{\star}$ & Sy1 & 207.8729 & -18.2297 & 0.012 & 42.34 & 42.03 & 42.62 & 20.18 & 1 \\
NGC5347 & Sy2  & 208.3243 & 33.4908 & 0.008 & 42.54 & 42.57 & 42.16 & >24.35 & 18 \\
CGCG074-129 & Sy2  & 212.6723 & 13.558 & 0.016 & 43.32 & 42.91 & 41.91 & 23.83 & This work \\
CircinusGalaxy$^{\star}$ & Sy2 & 213.2915 & -65.3392 & 0.001 & 43.38 & 42.64 & 42.53 & 24.8 & 19 \\
NGC5506$^{\star}$ & Sy2 & 213.3121 & -3.2076 & 0.006 & 43.3 & 43.2 & 42.93 & 22.44 & 1 \\
NGC5548$^{\star}$ & Sy1 & 214.4981 & 25.1368 & 0.017 & 43.52 & 43.32 & 43.15 & 20.69 & 1 \\
NGC5610$^{\star}$ & Sy2 & 216.0956 & 24.6141 & 0.017 & 43.38 & 43.06 & 42.72 & 22.56 & 1 \\
2MASXJ14320869-2704324 & Sy1 & 218.0363 & -27.0756 & 0.014 & 42.84 & 42.59 & - & - & - \\
WKK4438$^{\star}$ & Sy1 & 223.8225 & -51.5708 & 0.016 & 43.26 & 43.06 & 42.77 & 20.92 & 1 \\
IC4518A$^{\star}$ & Sy2 & 224.4216 & -43.1321 & 0.016 & 43.6 & 43.5 & 42.67 & 23.36 & 1 \\
NGC5861 & Sy2  & 227.317 & -11.3217 & 0.006 & 42.55 & 42.08 & - & - & - \\
NGC5990 & Sy2  & 236.5682 & 2.4154 & 0.012 & 43.58 & 43.29 & 41.00 & 23.43 & This work \\ 
IRAS15514-3729$^{s}$ & Sy2  & 238.6948 & -37.6387 & 0.019 & 43.18 & 42.82 & 42.38 & 22.78 & This work \\
WKK6092$^{\star}$ & Sy1 & 242.9642 & -60.6319 & 0.016 & 42.88 & 42.64 & 42.45 & 20 & 1 \\
ESO138-G001$^{\star}$ & Sy2 & 252.8339 & -59.2348 & 0.009 & 43.57 & 43.58 & 41.56 & 23.6 & 17 \\
ESO044-G007 & Sy2  & 258.98 & -73.3421 & 0.017 & 42.66 & 42.4 & - & - & - \\
NGC6300$^{\star}$ & Sy2 & 259.2478 & -62.8206 & 0.004 & 42.36 & 42.23 & 41.64 & 23.3 & 1 \\
ESO139-G012$^{\star}$ & Sy2 & 264.4128 & -59.9407 & 0.017 & 43.07 & 42.85 & 42.65 & 20 & 1 \\
IC4709$^{\star}$ & Sy2 & 276.0808 & -56.3692 & 0.017 & 43.14 & 42.92 & 43.07 & 23.15 & 1 \\
ESO103-G035$^{\star}$ & Sy2 & 279.5848 & -65.4276 & 0.013 & 43.79 & 43.71 & 43.41 & 23.28 & 1 \\
ESO140-G043$^{\star}$ & Sy1 & 281.2249 & -62.3648 & 0.014 & 43.64 & 43.67 & 43.13 & 20 & 1 \\
IC4769 & Sy2  & 281.9335 & -63.157 & 0.015 & 43.45 & 43.26 & 43.65 & 24.56 & This work \\
ESO281-G038$^{s}$ & Sy1 & 284.0874 & -43.1469 & 0.017 & 42.98 & 42.75 & 42.63 & 23.34 & This work \\ 
UGC11397$^{\star}$ & Sy2 & 285.9548 & 33.8447 & 0.015 & 42.88 & 42.63 & 42.63 & 22.87 & 1 \\
2MASXJ19373299-0613046$^{\star}$ & Sy1 & 294.3875 & -6.218 & 0.01 & 43.39 & 43.2 & 42.78 & 20.85 & 1 \\
NGC6860$^{\star}$ & Sy2 & 302.1954 & -61.1002 & 0.015 & 43.56 & 43.46 & 43.16 & 21.08 & 1 \\
2MASXJ20183871+4041003$^{\star}$ & Sy2 & 304.6613 & 40.6834 & 0.014 & 43.07 & 42.85 & 42.57 & 22.78 & 1 \\
IC4995  & Sy2 & 304.9957 & -52.622 & 0.016 & 43.25 & 43.05 & 43.11 &>24.68 & 16 , This work \\
MCG+04-48-002$^{\star}$ & Sy2 & 307.1461 & 25.7333 & 0.014 & 43.62 & 43.45 & 43.16 & 23.86 & 1 \\
ESO234-G050$^{\star}$ & Sy2 & 308.9912 & -50.1923 & 0.009 & 42.52 & 42.04 & 41.62 & 23.08 & 1 \\
IC5063$^{\star}$ & Sy2 & 313.0098 & -57.0688 & 0.011 & 43.85 & 43.8 & 43.1 & 23.56 & 1 \\
2MASXJ21025564+6336248 & AGN(?) & 315.7318 & 63.6069 & 0.011 & 42.59 & 41.66 & - & - & - \\
IC1368 & Sy2  & 318.5525 & 2.178 & 0.013 & 43.06 & 42.83 & - & - & - \\
IRAS21262+5643$^{\star}$ & Sy1 & 321.9391 & 56.943 & 0.014 & 43.51 & 43.33 & 43.1 & 20 & 1 \\
NGC7172$^{\star}$ & Sy2 & 330.5079 & -31.8697 & 0.009 & 42.98 & 42.8 & 42.67 & 22.91 & 1 \\
NGC7469$^{\star}$ & Sy1 & 345.8151 & 8.874 & 0.016 & 44.23 & 43.89 & 43.22 & 20.53 & 1 \\
NGC7479$^{\star}$ & Sy2 & 346.236 & 12.3229 & 0.008 & 43.08 & 43.2 & 41.89 & 24.76 & 12 \\
NGC7582$^{\star}$ & Sy2 & 349.5979 & -42.3706 & 0.005 & 43.3 & 42.81 & 43.48 & 24.23 & 12 \\
IC1490$^{s}$ & Sy1 & 359.7947 & -4.127 & 0.019 & 43.3 & 43.11 & 42.51 & 23.3 & This work  \\
\hline  

\end{longtable}
Notes: (I) Source name in \citep{asmus2020}. (II) Optical classification from SIMBAD or NED. (III) \& (IV) Optical coordinates (J2000)(V) Redshift (VI) Logarithm of W3 (12$\mu$m) continuum (egs/s) (VII) Logarithm of the nuclear 12$\mu$m luminosity of the AGN (egs/s)  (VIII) Logarithm of the intrinsic 2-10 keV (egs/s) (IX) Logarithm of the  column density along the line of sight ($\rm cm^{-2}$) (X) X-ray data reference: (1) \citet{ricci2017}, (2) \citet{Marchesi2018}, (3) \citet{guo2023}, (4) \citet{annuar2020}, (5) \citet{bauer2015}, (6) \citet{tanimoto2022}, (7) \citet{zhao2020}, (8) \citet{masini2016}, (9) \citet{marchesi2019}, (10) \citet{yamada2021}, (11) \citet{jiang2021}, (12) \citet{zhao2021}, (13) \citet{Boorman2016}, (14) \citet{Balokovic2014}, (15) \citet{lamassa2019} , (16) \citet{clavijo2022}, (17) \citet{sengupta2023}, (18) \citet{kammoun2019}, (19) \citet{arevalo2014}, \citet{Panagiotou2019}. \\
$^{\star}$ Sources detected in \swift/\swift 70 months all sky survey \\
$^{s}$ Sources with spectra from \swift/XRT \\
\end{landscape}

\end{appendix}

\end{document}